\documentstyle[aps,eqsecnum,epsf]{revtex}

\input{epsf}
\newcommand{\be}{\begin{eqnarray}}
\newcommand{\ee}{\end{eqnarray}}
\newcommand{\beq}{\begin{equation}}
\newcommand{\eeq}{\end{equation}}

\newcommand{\nablaright}{\stackrel{\rightarrow}{\nabla}}
\newcommand{\nablaleft}{\stackrel{\leftarrow}{\nabla}}
\newcommand{\nablaboth}[1]{\stackrel{\leftrightarrow}{\nabla}{}_{\!\! #1}}
\begin{document}
\begin{flushright}
JLAB-THY-00-37 \\
RUB-TPII-17/00 \\
hep-ph/0010296
\end{flushright}
\vspace{1cm}
\begin{center}
{\bf\Large DVCS amplitude at tree level: \\[.15cm] 
Transversality, twist--3, and factorization} \\[1cm]
\end{center}
\begin{center}
{A.V. RADYUSHKIN$^{a,b,1}$, C. WEISS$^{c,2}$} 
\\[2mm]
{\em $^a$Physics Department, Old Dominion University,} \\
{\em Norfolk, VA 23529, USA}
\\[2mm] 
{\em $^b$Theory Group, Jefferson Lab,} \\
{\em Newport News, VA 23606, USA}
\\[2mm] 
{\em $^c$ Institut f\"ur Theoretische Physik II} \\
{\em Ruhr--Universit{\"a}t, D--44780 Bochum, Germany}
\end{center}
\vspace{1cm}
\begin{abstract}
We study the virtual Compton amplitude in the generalized Bjorken 
region ($q^2 \rightarrow \infty$, $t$ small) in QCD by means of a
light--cone expansion of the product of e.m.\ currents in string 
operators in coordinate space. Electromagnetic gauge invariance 
(transversality) is maintained by including in addition to the 
twist--2 operators ``kinematical'' twist--3 operators which appear 
as total derivatives of twist--2 operators. 
The non-forward matrix elements of the elementary twist--2 operators are 
parametrized in terms of two-variable spectral functions 
(double distributions), from which twist--2 and 3 skewed distributions
are obtained through reduction formulas. Our approach is equivalent to a 
Wandzura--Wilczek--type approximation for the twist--3 skewed distributions. 
The resulting Compton amplitude is manifestly transverse up to terms of 
order $t/q^2$. We find that in this approximation the tensor amplitude for 
longitudinal polarization of the virtual photon is finite, while the one
for transverse polarization contains a divergence already 
at tree level. However, this divergence has zero projection on the
polarization vector of the final photon, so that the physical helicity 
amplitudes are finite.
\end{abstract}
\vfill
\rule{5cm}{.2mm} \\
{\footnotesize $^{\rm 1}$ E-mail: radyush@jlab.org\ . Also at Laboratory 
of Theoretical Physics, JINR, Dubna, Russia.} \\
{\footnotesize $^{\rm 2}$ E-mail: weiss@tp2.ruhr-uni-bochum.de} \\
\newpage
\tableofcontents
\newpage
\section{Introduction}
Deeply virtual Compton scattering (DVCS) is currently receiving a lot of
attention, as an exclusive process closely related to inclusive
deep--inelastic scattering (DIS)
\cite{Ji:1997nm,Collins:1997fb,Radyushkin:1997ki}. The theory of DIS is
well understood and can be formulated both in the field--theoretical
framework of QCD or the particle--based language of the parton model. The
relevant information about the non-perturbative structure of the nucleon is
contained in expectation values of QCD light--ray operators in the nucleon,
which can be interpreted as the parton distributions. Similarly, the DVCS
amplitude in the asymptotic region can be expressed through off--diagonal
(non-forward) matrix elements of light--ray operators, which also possess a
partonic interpretation in terms of so-called ``skewed'' parton
distributions (SPD's). The initial motivation for considering DVCS was the
hope that by measuring the SPD's one could obtain information about the
angular momentum distributions of quarks and gluons in the nucleon
\cite{Ji:1997nm}. However, it was soon realized that the DVCS process is of
interest in itself, independently of its connection to the nucleon spin
problem. It is the simplest representative of a whole class of exclusive
electroproduction processes, including also hard meson production, for
which factorization can be proven in QCD at leading twist level
\cite{Collins:1997fb,Radyushkin:1997ki}.  The first experimental
observations of DVCS have recently been reported by the ZEUS \cite{ZEUS},
H1 \cite{H1}, and HERMES \cite{HERMES} collaborations, and an extensive
experimental program is planned at JLAB \cite{JLAB}. The DVCS process is
also related, by crossing relations \cite{Polyakov:1999ze,Polyakov:1999gs},
to the electroproduction of pairs of hadrons off a real photon, such as
$\gamma^* \gamma \rightarrow 2 \pi$ \cite{Diehl:1998dk}, which can be
studied at $e^+ e^-$ colliders.
\par
The virtual Compton amplitude in the asymptotic region of large virtuality
of the incoming photon, $-q_1^2 \rightarrow \infty$, and small momentum
transfer squared, $t \ll 1 \,{\rm GeV}^2$, has been studied in QCD using
expansion techniques familiar from DIS: The collinear expansion in momentum
space \cite{Ellis:1983cd}, and the non-local light--cone expansion in
coordinate space \cite{Anikin:1978tj,Balitsky:1989bk}. Originally only the
contribution from twist--2 operators was included in the calculation of the
tensor amplitude \cite{Ji:1997nm,Radyushkin:1997ki}. This approximation
does not include subleading terms proportional to the transverse part of
the momentum transfer, $r_{\perp} \equiv \Delta$, and the tensor amplitude
thus obtained formally violates electromagnetic gauge invariance
(transversality) by terms linear in $\Delta$.\footnote{Bl\"umlein and
Robaschik \cite{Blumlein:2000cx} studied instead of the tensor DVCS
amplitude the individual photon--hadron helicity amplitudes. The problem
with the gauge invariance of the twist--2 contribution does not arise as
long as one considers only the leading helicity amplitudes in
$1/\sqrt{-q_1^2}$.}  To remedy this, Guichon and Vanderhaeghen
\cite{Guichon} proposed to add a term linear in $\Delta$, producing a
tensor amplitude which is transverse up to terms of order $\Delta^2 \sim
t$. It is clear that in a QCD treatment such a term could come only from
contributions of operators of higher twist ($>2$).  Indeed, calculations by
various groups
\cite{Anikin:2000em,Penttinen:2000dg,Belitsky:2000vx,Radyushkin:2000jy}
have recently shown that the term introduced in Ref.\cite{Guichon} appears
naturally as part of the twist--3 contribution to the Compton amplitude.
Anikin {\it et al.} \cite{Anikin:2000em}, applying the momentum--space
collinear expansion up to twist--3 level, reproduced not only the
``improved'' twist--2 structure of Ref.\cite{Guichon}, but found also
other, separately transverse twist--3 structures in the tensor amplitude,
which can contribute to the observables of the DVCS process. A similar
result was obtained by Penttinen {\it et al.} \cite{Penttinen:2000dg}, who
computed the DVCS amplitude to accuracy $1/\sqrt{-q_1^2}$ in a parton model
approach.  Within the framework of the coordinate--space light--cone
expansion the twist--3 contribution to the DVCS amplitude was studied by
Belitsky and M{\"u}ller \cite{Belitsky:2000vx}. In particular, these
authors demonstrated that in order to get a gauge invariant result up to
terms of order $t/q_1^2$ it is sufficient to retain only a certain part of
the full twist--3 SPD's which is obtained by Wandzura--Wilczek--type
relations from the twist--2 distributions.  Kivel {\it et al.}\
\cite{Kivel:2000rb} noted that the Wandzura--Wilczek expression for the tensor 
amplitude contains a divergence in the part corresponding to transverse polarization
of the virtual photon.
\par
In this paper we address the question of transversality of the DVCS
amplitude within the framework of the light--cone expansion of the current
product in QCD string operators in coordinate space developed by Balitsky
and Braun \cite{Balitsky:1989bk}.  (A short summary of our results has
already been given in Ref.\cite{Radyushkin:2000jy}.) The investigation
consists of two main parts. First, we show that transversality of the
light--cone expansion can be achieved, in the minimal sense, by including
in the expansion certain operators which are algebraically of twist 3 but
given by total derivatives of twist--2 string operators.  We shall refer to
these as ``kinematical'' twist--3 contributions, in order to distinguish
them from the twist--3 contributions involving quark--gluon operators which
cannot be reduced to total derivatives by the equations of motion. The
operations necessary to include the ``kinematical'' twist--3 contributions
in the light--cone expansion can be performed using the calculus of QCD
string operators. The result has the same form as the usual twist--2 part
of expansion, but with certain finite translations of the center
coordinates of the twist--2 string operators (similar expressions were
derived independently in Refs.\cite{Belitsky:2000vx,Kivel:2000rb}).
\par
Second, we investigate whether the approximation of keeping only the
``kinematical'' twist--3 contributions required by transversality leads to
useful results at the level of the hadronic DVCS amplitude. To this end, we
parametrize the matrix elements of the fundamental twist--2 operators in
terms of two--variable spectral functions (double distributions)
\cite{Radyushkin:1997ki,Radyushkin:1999es,Radyushkin:1999bz,Polyakov:1999gs}.
The Compton amplitude in the DVCS limit, including the ``kinematical''
twist--3 contributions, can be expressed in terms a number of one-variable
skewed distributions (SPD's), which are obtained in the form of
one--dimensional reductions of the twist--2 double distribution.  The
resulting tensor amplitude is manifestly transverse up to terms of order
$t/q_1^2$ or $m^2/q_1^2$ ($m$ is the target mass), {\it i.e.}, its
contraction with the four--momentum of the initial photon, 
$q_{1\mu} T_{\mu\nu}$, and that of the final photon, $T_{\mu\nu} q_{2\nu}$,
are both of order $t/q_1^2$ or $m^2/q_1^2$. We find that our approximation
gives a finite result for the part of the tensor amplitude corresponding to
longitudinal polarization of the initial photon, while the part
corresponding to transverse polarization contains a divergence. However, as
we shall show, the divergent part has zero projection on the polarization
vector of the final (real) photon, so it does not affect the physical
helicity amplitudes.
\par
Our approach of keeping only the ``kinematical'' twist--3 operators in the
coordinate--space light cone expansion is actually equivalent to the
Wandzura--Wilczek--type approximation for the twist--3 SPD's introduced in
Ref.~\cite{Belitsky:2000vx}. In addition to giving an alternative
derivation of this approximation, an important advantage of our treatment
is the possibility to incorporate the formalism of double distributions
\cite{Radyushkin:1997ki,Radyushkin:1999es,Radyushkin:1999bz,Polyakov:1999gs}.
The twist--2 as well as the new twist--3 SPD's are obtained by reduction
formulas from the fundamental twist--2 double distribution. This allows for
a straightforward derivation of the properties of these new functions and
the Wandzura--Wilczek--type relations between them. In particular, the fact
that this approximation gives finite results for the tensor amplitude for
longitudinally polarized initial photon, but contains a divergence in the case of
transverse polarization (which, however, drops out of the physical helicity
amplitudes), can immediately be understood on the basis of the reduction
formulas.
\par
The plan of this paper is as follows. Section~\ref{sec_virtual} gives a
brief summary of the kinematics of virtual Compton scattering in the
generalized Bjorken limit.  In Section~\ref{sec_light} we discuss the
peculiarities of the light--cone expansion in the ``non-forward'' case,
where total derivative operators contribute. Our treatment is based on the
formalism developed by Balitsky and Braun \cite{Balitsky:1989bk}. In
Section~\ref{subsec_twist} we recall the main steps in the twist
decomposition of string operators. In Section~\ref{subsec_transversality}
we investigate the problem of transversality of the light--cone expansion
at operator level in coordinate space. We observe that the twist--2
contribution is not transverse, and show that transversality can be
restored by including total derivatives of twist--2 operators, which are
algebraically of twist 3. We first formulate this as as an iterative
procedure generating an infinite series of total derivatives of twist--2
operators, then show that this series can be summed up in closed form to
give finite translation operators acting on the twist--2 string operators
(details are given in Appendix~\ref{app_deconstructing}).  A manifestly
transverse expression for the light--cone expansion of the current product
in terms of twist--2 operators and their total translations is presented.
In Section~\ref{sec_amplitude} we use our result for the light--cone
expansion including ``kinematical'' twist--3 terms to compute the hadronic
DVCS amplitude. For simplicity we consider the case of a pion target, which
has spin 0 and definite $C$--parity.  In Section~\ref{subsec_spectral} we
describe the spectral representation of the fundamental twist--2 matrix
elements in terms of double distributions.  The matrix elements of the
vector--type string operators including the ``kinematical'' twist--3 terms
are then derived in Section~\ref{subsec_vector}.  The new SPD's are
introduced through reduction formulas, and their properties are
discussed. The DVCS amplitude for the pion is computed in
Section~\ref{subsec_amplitude}. We inspect the singular integrals resulting
from the inclusion of the ``kinematical'' twist--3 terms. We show that the
tensor amplitude for transverse polarization of the initial (virtual)
photon contains a divergence, but that this divergence is ``harmless'' in
the sense that it has zero projection on the polarization vector of the
final photon.  In Section~\ref{subsec_WW} we reformulate our results for
the matrix elements and the Compton amplitude, introducing a new SPD as a
Wandzura--Wilczek--type transformation of the basic twist--2 SPD.  This
allows us to demonstrate that our approach is equivalent to the
Wandzura--Wilczek approximation proposed in Ref.\cite{Belitsky:2000vx}. In
this language, the singularity for transverse polarization described in
Section~\ref{subsec_amplitude} appears due to a discontinuity of the
Wandzura--Wilczek--transformed SPD at $x = \xi$, which occurs
independently of the dynamical behavior of the underlying twist--2 matrix
elements.  Our conclusions are summarized in Section~\ref{sec_conclusions}.
\par
In this paper we study the properties of the DVCS amplitude including
``kinematical'' twist--3 contributions at tree level. We shall not be
concerned with logarithmic corrections resulting from the scale dependence
of the operators (the evolution of the skewed distributions).  For the
twist--2 operators this problem has exhaustively been treated in the
literature, see {\it e.g.}\
Refs.\cite{Ji:1997nm,Radyushkin:1997ki,Bartels:1982jh,Geyer:1985vw,Dittes:1988xz,Frankfurt:1998ha,Belitsky:1998pc}. The
generalization of the evolution equations to the ``kinematical'' twist--3
contribution discussed in the present paper is in principle straightforward
but will not be pursued here.
\section{Virtual Compton amplitude in the generalized Bjor\-ken limit}
\label{sec_virtual}
{\it Virtual Compton scattering.} For a unified description of the 
kinematics of processes such as DIS and DVCS let us imagine a general 
virtual Compton scattering process off a hadron,
\beq
\gamma^\ast (q_1 ) \; + \; h (p_1 ) \; \rightarrow \;
\gamma^\ast (q_2 ) \; + \; h (p_2 ) ,
\label{VCS}
\eeq
where the incoming photon has space--like virtuality, $q_1^2 < 0$, and
the final photon may have any virtuality allowed by four--momentum 
conservation; in particular, we shall allow for the final photon to be 
real, $q_2^2 = 0$. It is convenient to use as independent momentum variables
the average of the photon and hadron momenta, $q$ and $p$,
and the momentum transfer, $r$:
\beq
p_{1, 2} \;\; = \;\; p \pm \frac{r}{2} ,
\hspace{5em}
q_{1, 2} \;\; = \;\; q \mp \frac{r}{2} .
\label{q_p_r_def}
\eeq
From the mass shell conditions, $p_1^2 = p_2^2 = m^2$, it 
follows that
\beq
(p r) \;\; = \;\; 0, \hspace{5em} p^2 \;\; = \;\; m^2 - \frac{t}{4} ,
\hspace{5em} t \;\; \equiv \;\; r^2 .
\eeq
In the following we shall neglect the target mass and put $m = 0$.
\par
The amplitude for the virtual Compton process is defined by the
transition matrix element of the time--ordered product of two
electromagnetic current operators between the hadronic states:
\be
T_{\mu\nu} \;\; = \;\; i \int d^4 z \; e^{i (q z)}
\langle p - r/2 | {\rm T}\, J_\mu (-z/2) J_\nu (z/2) 
| p + r/2 \rangle .
\label{Compton_def}
\ee
This tensor is transverse with respect to the incoming and 
outgoing photon momenta:
\beq
\left( q - \frac{r}{2} \right)_{\mu} T_{\mu\nu} \;\; = \;\; 0 , 
\hspace{5em}
T_{\mu\nu} \left( q + \frac{r}{2} \right)_{\nu} \;\; = \;\; 0 ,
\label{transversality}
\eeq
which is an immediate consequence of current conservation, 
$\partial_\mu J_\mu (x) = 0$.\footnote{Strictly speaking this 
is correct only up to terms arising from the differentiation of the
step functions in time occurring in the reduction of the 
time-ordered product (``seagull terms'').
We shall not consider these contributions, since they do not play
a role at large $q^2$.} Separating the symmetric and antisymmetric
tensor parts of $T_{\mu\nu}$ (with otherwise same arguments),
\beq
\left. \begin{array}{c} T_{\{\mu\nu\}} \\[1.5ex]
T_{[\mu\nu ]} \end{array}  \right\} 
\;\; \equiv \;\; 
\frac{1}{2} \left( T_{\mu\nu} \pm T_{\nu\mu} \right) ,
\eeq
the two conditions Eq.(\ref{transversality}) can also be stated
in the form
\be
q_{\mu} T_{\{ \mu\nu \}} &=& \frac{r_{\mu}}{2} 
T_{[\mu\nu ]} , \\
q_{\mu} T_{[\mu\nu ]} &=& 
{r_{\mu}\over 2} T_{\{\mu\nu\}} .
\label{trans} 
\ee
One notes that in the case of non-forward
scattering, $r \neq 0$, the transversality conditions generally imply
relations between the symmetric and antisymmetric parts of the amplitude.
It is only in the limit of forward scattering, $r = 0$, (as is relevant
{\it e.g.}\ to deep--inelastic scattering) that the two transversality
conditions decouple and turn into conditions on the symmetric
and the antisymmetric parts separately.
\par
{\it Generalized Bjorken region: DIS and DVCS.}
We shall consider the amplitude for the virtual Compton process,
Eq.(\ref{VCS}), in two different asymptotic regions, both
characterized by large virtuality of the incoming photon. One is the
region corresponding to deep--inelastic scattering (DIS), where one is
interested in the forward scattering amplitude 
($q_1 = q_2 = q, p_1 = p_2 = p, r = 0$) in the limit 
$q_1^2 \rightarrow \infty, (p_1 q_1) \rightarrow \infty$, with
$x_B \equiv -q_1^2 / [2 (p_1 q_1)]$ fixed. The other is the
region of deeply virtual Compton scattering (DVCS), where one is
dealing with the amplitude for production of a real photon, 
$q_2^2 = 0$, in the limit 
$q_1^2 \rightarrow \infty, (p_1 q_1) \rightarrow \infty$, again with
$x_B \equiv -q_1^2 / [2 (p_1 q_1)]$ fixed. This process requires
a non-zero momentum transfer from the initial to the final hadron,
($p_1 - p_2 \equiv r \neq 0$), with $(r q_1) \rightarrow \infty$
proportional to $q_1^2$. One considers a situation where 
$t \equiv r^2 < 0$ does not grow with $q_1^2$. In order to treat both
asymptotic regions within a common framework it is convenient to
introduce two scaling variables,
\beq
\xi \;\; \equiv \;\; \frac{-q^2}{2 (p q)} , 
\hspace{5em}
\eta \;\; \equiv \;\; \frac{(r q)}{2 (p q)} ,
\label{xi_eta_def}
\eeq
where $\eta$ is referred to as ``skewedness''.
The virtualities of the incoming and outgoing photon can then 
be expressed as
\be
q_1^2 &=& \left( 1 + \frac{\eta}{\xi} \right) q^2 + \frac{t}{4}
\label{q_1_squared_from_xi_eta} ,
\\
q_2^2 &=& \left( 1 - \frac{\eta}{\xi} \right) q^2 + \frac{t}{4}
\label{q_2_squared_from_xi_eta} ,
\ee
and the Bjorken variable becomes
\beq
x_B \;\; = \;\; \frac{-q_1^2}{2 (p_1 q_1)}
\;\; \equiv \;\; \frac{\xi + \eta}{1 + \eta} .
\eeq
Because of Eq.(\ref{q_1_squared_from_xi_eta}) we may generally use $q^2$ 
instead of $q_1^2$ as the large invariant. The two situations, DIS and
DVCS, now correspond to\footnote{Because of the identification 
of $\xi$ with $\eta$ in DVCS kinematics, $\xi$ in DVCS is frequently also
referred to as ``skewedness''.}
\beq
\begin{array}{rll} 
\mbox{DIS:} & \;\; \eta \; = \; 0, & \;\; x_B \; = \; \xi ,\\[2ex]
\mbox{DVCS:} & \;\; \eta \; = \; \xi, & \;\; x_B \; = \; 
\displaystyle{\frac{2 \xi}{1 + \xi}} .
\end{array}
\eeq
\par
{\it Aim of the present calculation.} In this paper we calculate
the virtual Compton amplitude, Eq.(\ref{Compton_def}), in DVCS kinematics, 
neglecting contributions of order $t/q^2$ or $m^2/q^2$. Specifically, we aim 
to compute the tensor amplitude to an accuracy such that the transversality 
conditions, 
Eqs.(\ref{transversality}) {\it viz.} (\ref{trans}), 
are satisfied up to terms of order $t/q^2$ or $m^2/q^2$. In this 
approximation the kinematical limits for $\xi$ are
\beq
0 \leq \xi \leq 1 ,
\eeq
which is equivalent to $0 \leq x_B \leq 1$. 
For DVCS it is convenient to separate from the momentum transfer, $r$,
the component parallel to $p$, which is 
responsible for producing the large scalar when contracting with $q$, 
{\it i.e.}, to write
\beq
r \;\; = \;\; 2 \xi p \; + \; \Delta .
\label{r_thru_Delta}
\eeq
The remaining part of the momentum transfer, $\Delta$, satisfies
$(\Delta q) = -t/4$ and $(\Delta p) = -\xi t/2$. 
Since $p$ is time-like and $q$ space-like, and since furthermore
$\Delta$ itself is space--like, $\Delta^2 = (1 - \xi^2 ) t < 0$,
it follows that all components of $\Delta$ must be of order $|t|^{1/2}$:
\beq
\Delta \;\; \propto \;\; |t|^{1/2} .
\eeq
This means that when computing the DVCS amplitude using the light--cone
expansion we may expand the ingredients --- the  momentum--space 
quark propagators and
the hadronic matrix elements --- in $\Delta$, since this four--vector 
cannot produce any large scalars even when contracted with the 
photon momentum, $q$. 
This would not be possible for the full momentum transfer, $r$, since 
$(rq)$ is not small. In the language of the parton model, 
the component $2 \xi p$ of the momentum transfer would be identified with
the longitudinal component, while $\Delta$ would be the 
transverse component, $r_\perp$. Note that in our approach the 
decomposition of the momentum transfer, Eq.(\ref{r_thru_Delta}), 
appears on purely kinematical grounds, without reference to a 
particular frame.
\section{Light--cone expansion in the non-forward case}
\label{sec_light}
\subsection{Twist decomposition of string operators: Twist 2}
\label{subsec_twist}
The behavior of the virtual Compton amplitude, Eq.(\ref{Compton_def}),
in the generalized Bjorken limit (DIS or DVCS) is dominated by the 
light--cone singularity of the time--ordered product of the two 
electromagnetic currents, which is a consequence of the asymptotic
freedom of QCD. An asymptotic expansion of the Compton amplitude
in inverse powers of $1/q^2$ can be derived from the light--cone
expansion of the current product. In this section we discuss 
the peculiarities of the light--cone expansion in the ``non-forward'' 
case, {\it i.e.}, finite momentum transfer in the
matrix elements, as is relevant {\it e.g.}\ for DVCS. 
An additional complication compared
to the ``forward'' case arises due to the fact that total derivatives of 
operators can contribute to the matrix elements, which greatly enlarges
the set of operators participating in the expansion to a given accuracy.
In particular, we show here that keeping the total derivatives of 
twist--2 operators (which are operators formally of twist 3)
is essential for maintaining transversality of
the leading term in the light--cone expansion. We shall proceed step by
step. We first investigate the well--known twist--2 contribution and 
show that, in the non-forward case, it is not transverse to the accuracy
required. We then formulate a procedure how transversality can be 
``repaired'' by including total derivatives of twist--2 operators,
order--by--order in the number of total derivatives. Finally, 
we show that the series 
of total derivatives of twist--2 operators can be summed up in closed 
form, resulting in an expression involving certain finite translations
of the original twist--2 operators, which is manifestly transverse.
\par
{\it Expansion in string operators.} We shall be concerned 
with the light--cone expansion of the current product,
\beq
\Pi_{\mu\nu} (x, y) \;\; \equiv \;\; i {\rm T} J_{\mu}(x) J_{\nu} (y) ,
\label{Pi}
\eeq
where $J_\mu (x)$ is the vector current operator. 
It is convenient to introduce the center and relative 
coordinates of the two points, $x$ and $y$,
\beq
X \;\; = \;\; (x + y)/2 , 
\hspace{5em} z \;\; = \;\; y - x ,
\hspace{5em}
\left. \begin{array}{c} x \\ y \end{array} \right\}
\;\; = \;\; X \mp z/2 ,
\label{COM_relative_coordinates}
\eeq
and to regard the current product as a function of these variables.  The
light--cone expansion is derived from a formal expansion of the current
product in QCD string (non-local) operators in a background gauge field,
which are subsequently decomposed in operators of definite twist. This
expansion is generated by contracting the quark fields in Eq.(\ref{Pi})
with the quark Green function in a background gauge field, and expanding in
powers of insertions of the background field in the Fock--Schwinger
gauge. The resulting expansion can be formulated in a gauge--invariant way
by introducing QCD string operators with fields connected by gauge
links.  It is known that the leading light--cone singularity is contained
in the contraction corresponding to the simplest two--point string operator
(the ``handbag'' diagram):
\be
\Pi_{\mu\nu} (z| X) &\equiv& i {\rm T} J_{\mu}(X - z/2) J_{\nu} (X + z/2) 
\nonumber 
\\[1.5ex] 
&=& \bar\psi (X - z/2) \frac{\gamma_\mu \hat z \gamma_\nu}{2\pi^2 z^4}
[X - z/2, X + z/2] \psi (X + z/2) 
\nonumber \\[1.5ex] 
&+& (z \rightarrow -z, \mu \leftrightarrow \nu) .
\nonumber \\[1.5ex] 
&+& \mbox{terms $1/z^2, \ldots$}
\label{handbag_three_gamma}
\ee
Here
\beq
\frac{\hat z}{2 \pi^2 (z^2 - i 0)^2} \;\; = \;\; \int\frac{d^4 k}{(2\pi )^4}
e^{-i (kz)} \frac{\hat k}{k^2 + i 0} ,
\hspace{5em} \hat z \;\; \equiv \;\; z_\mu \gamma_\mu \;\;\; \mbox{\it etc.},
\label{propagator_z}
\eeq
is the free quark propagator in coordinate representation.
Reducing the product of three gamma matrices, Eq.(\ref{handbag_three_gamma}) 
can be written as
\be
\lefteqn{ \Pi_{\mu\nu} (z| X) } && \nonumber \\
&=& \frac{2 z_\rho}{\pi^2 z^4} \left\{ \phantom{-i} s_{\mu\rho\nu\sigma} 
\left[ \bar\psi (X-z/2) \gamma_\sigma [X-z/2, X+z/2] 
\psi (X+z/2) \; - \; (z \rightarrow -z) \right] 
\phantom{\frac{2 z_\rho}{\pi^2 z^4}}
\right.
\nonumber \\
&& \left. 
\phantom{\frac{z_\rho}{2\pi^2 z^4}}
- i \epsilon_{\mu\rho\nu\sigma} 
\left[ \bar\psi (X - z/2) \gamma_\sigma \gamma_5 [X - z/2, X + z/2] 
\psi (X + z/2) \; + \; (z \rightarrow -z) \right]
\right\} ,
\label{handbag_string}
\ee
where
\beq
s_{\mu\rho\nu\sigma} \;\; = \;\;
\frac{1}{4} {\rm tr} \left[ \gamma_\mu \gamma_\rho \gamma_\nu 
\gamma_\sigma \right]
\nonumber \\
\;\; = \;\; g_{\mu\rho} g_{\nu\sigma} - g_{\mu\nu} g_{\rho\sigma}
+ g_{\mu\sigma} g_{\nu\rho} .
\eeq
In the string operators in Eq.(\ref{handbag_three_gamma}) and
(\ref{handbag_string}) the gauge link runs along the 
straight line connecting the two space--time points:
\beq
[x, y] \;\; \equiv \;\; {\rm P} \exp i \int_0^1 du \; (x - y)_\alpha
A_\alpha (u x + \bar u y), 
\hspace{5em} \bar u \equiv 1 - u .
\label{link_x_y}
\eeq
For brevity we shall in the following not write the links 
between the fields; it will always be understood that the fields
in the string operator (including possible insertions of gauge fields at 
intermediate points) are connected by gauge links along the straight line.
\par
{\it Differentiating string operators.}
For a given background gauge field the link operator defined by 
Eq.(\ref{link_x_y}) is a uniquely defined function of the end points.
In particular, when differentiating the string operator with respect to
its end points one is changing the contour of integration by an
infinitesimal amount, so the link operator gets differentiated as well. 
There are generally two contributions from differentiating the link
operators. One gives the gauge potentials at the end points, which combine
with the derivatives acting on the quark fields to give covariant 
derivatives. The other can be expressed as a line integral of the 
gauge field along the contour of the string. For the derivative of
the string operators in Eq.(\ref{handbag_string}) with respect to the 
relative coordinate we find:
\be
\lefteqn{
\frac{\partial}{\partial z_\mu} \bar\psi (X - z/2) 
\gamma_\sigma \psi (X + z/2) }
&& \nonumber \\
&=& \frac{1}{2} \, \bar\psi (X - z/2) 
\left( \nablaleft_\mu \gamma_\sigma 
\; + \; \gamma_\sigma \nablaright_\mu \right) \psi (X + z/2)
\nonumber \\
&-& i \int_{-1}^1 dt\, t\; \bar\psi (X - z/2) \gamma_\sigma z_\alpha 
F_{\mu\alpha} (X + tz/2) \psi (X + z/2) ,
\label{string_derivative_z}
\ee
where gauge links between the fields are implicit everywhere; for the string
operator with  Dirac matrix $\gamma_\sigma \gamma_5$ one should replace
$\gamma_\sigma \rightarrow \gamma_\sigma \gamma_5$ everywhere.
The derivative with respect to the center coordinate (we refer to it
as ``total'' derivative) takes the form
\be
\lefteqn{
\frac{\partial}{\partial X_\mu} \bar\psi (X - z/2) 
\gamma_\sigma \psi (X + z/2) }
&& \nonumber \\
&=& \bar\psi (X - z/2) \left( - \nablaleft_\mu \gamma_\sigma 
\; + \; \gamma_\sigma \nablaright_\mu \right)
\psi (X + z/2)
\nonumber \\
&-& 2 i \int_{-1}^1 dt\; \bar\psi (X - z/2) \gamma_\sigma z_\alpha 
F_{\mu\alpha} (X + tz/2) \psi (X + z/2) .
\label{string_derivative_X}
\ee
Here
\be
\nablaright_\mu \psi (x) &\equiv& \phantom{-} 
\partial_\mu \psi (x) - i A_\mu (x) 
\psi (x) , 
\\
\bar\psi (x) \nablaleft_\mu &\equiv& -\partial_\mu \bar \psi (x) 
- i \bar \psi (x) A_\mu (x)
\ee
are the right and left covariant derivatives\footnote{Since we are dealing
with derivatives of fields of modified arguments it is important to maintain 
the distinction between the gradient evaluated at the point $z/2$,
$\partial_\mu \psi (z/2) \equiv (\partial / \partial x_\mu ) \psi 
(x)_{x = z/2}$, and the derivative 
$(\partial / \partial z_\mu) \psi (z/2 ) = (1/2) \partial_\mu \psi (z/2)$.},
and 
\beq
F_{\mu\alpha} \;\; \equiv \;\; 
i \left[ \nablaright_\mu , \nablaright_\alpha \right]
\eeq
the field strength.
\par
{\it Twist--2 part of string operators.}
The string operators in Eq.(\ref{handbag_string}) have as yet no definite
twist. Consequently, the ``handbag'' contribution to the current product,
Eq.(\ref{handbag_string}), as it stands, contains also subleading
contributions in $1/z^2$, which would be of the same order ($1/z^2$ or less
singular) as contributions from quark--gluon string operators not included in
Eq.(\ref{handbag_string}).  The leading $1/z^4$ singularity is given by the
twist--2 part of the string operators. The latter is defined as those
operators obtained by formally Taylor--expanding the string operators in
Eq.(\ref{handbag_string}) in the separation, $z$, and retaining only the
totally symmetric traceless parts of the coefficients in the
expansion:\footnote{Note that this definition is at a purely algebraic
level and does not imply any statement about the convergence of the series
of local operators.}
\be
\lefteqn{
\left[ \bar\psi (X - z/2) \gamma_\sigma \psi (X + z/2) \right]^{\rm twist-2}
} && \nonumber \\
&\equiv& 
\sum_{n = 0}^\infty \frac{1}{n!} \;
z_{\alpha_1} \ldots z_{\alpha_n} \;
\bar\psi (X) \left[
\gamma_{\left\{ \sigma \right. } \nablaboth{\alpha_1}
\ldots \nablaboth{\left. \alpha_n \right\}} 
- \mbox{traces} \right] \psi (X) , 
\label{gamma_LT}
\ee
and similarly for the operator with Dirac matrix $\gamma_\sigma \gamma_5$.
Here $\nablaboth \equiv (\nablaright + \nablaleft )/2$. 
As was shown in Ref.\cite{Balitsky:1989bk}, the
two operations, ``symmetrization'' and ``subtraction of traces'', 
can in fact be carried out directly at the level of non-local operators,
without having to go through the series of local operators \cite{Geyer2000}.
The part of the string operator corresponding to totally symmetric
local tensor operators is projected out by the operation
\beq
\left[ \bar\psi (X - z/2) \gamma_\sigma \psi (X + z/2) \right]^{\rm sym}
\;\; = \;\; 
\frac{\partial}{\partial z_\sigma} \int_0^1 dt \; 
\bar\psi (X - t z/2) \hat{z} \psi (X + t z/2) .
\label{string_sym}
\eeq
The subtraction of traces can be achieved by noting that the tracelessness 
of the local operators in the expansion, Eq.(\ref{gamma_LT}), implies that 
the string operator contracted with $z_\sigma$ should satisfy the d'Alembert 
equation with respect to $z$:
\beq
\Box_z \; \left[ \bar\psi (X - z/2) \hat{z} \psi (X + z/2) 
\right]^{\rm traceless}
\;\; = \;\; 0 .
\label{harmonic_O}
\eeq
Thus, combining the two operations, the twist--2 part of the vector string 
operator (for arbitrary $z^2 \neq 0$) is obtained
by substituting in the R.H.S.\ of Eq.(\ref{string_sym}) the solution of 
Eq.(\ref{harmonic_O}) with ``initial conditions'' given by the operator 
$\bar\psi (X - z/2) \hat z \psi(X + z/2)$ at $z^2 = 0$. On the light cone ($z^2 = 0$) 
trace subtraction is irrelevant, and the twist--2 part of the vector string operator
coincides with the symmetric part as defined by Eq.(\ref{string_sym}).
\par
When parametrizing the hadronic matrix elements of the
operators, the $z^2$-- (``trace'') terms in the matrix element are
proportional to dimensionful scalars characterizing the target
--- the target mass, $m^2$, or the square of the momentum transfer, $t$.
The inclusion of the $z^2$ terms in the calculation of the 
Compton amplitude then generates $m^2/q^2$-- (target mass)  and 
$t/q^2$--corrections to the DVCS amplitude. Here we shall neglect such 
corrections, so that the $z^2$--terms in the matrix elements need not
be subtracted explicitly.
\par
{\it Twist--2 contribution to the light--cone expansion.}
The twist--2 part of the current product, which we denote by 
$\Pi_{\mu\nu}^{(0)}$, is given by 
Eq.(\ref{handbag_string}), with the string operators replaced by 
their symmetric parts (we do not worry about trace subtraction):
\be
\Pi_{\mu\nu}^{\rm (0)} (z|X)
&=& \frac{2 z_\rho}{\pi^2 z^4} \left\{ \phantom{-i}
s_{\mu\rho\nu\sigma} 
\left[\bar\psi (X - z/2) \gamma_\sigma \psi (X + z/2) 
\phantom{\gamma_5} - (z \rightarrow -z)
\right]^{\rm sym}  \right.
\nonumber \\
&& \phantom{\frac{z_\rho}{2\pi^2 z^4}} \left. 
-i \epsilon_{\mu\rho\nu\sigma} 
\left[\bar\psi (X - z/2) \gamma_\sigma \gamma_5 
\psi (X + z/2) + (z \rightarrow -z)
\right]^{\rm sym} 
\right\} .
\label{handbag_twist_2}
\ee
Using the property Eq.(\ref{harmonic_O}), and the fact that the
coefficient function --- the free fermion propagator, Eq.(\ref{propagator_z})
--- satisfies
\beq
\frac{\partial}{\partial z_\rho} \left( \frac{z_\rho}{2\pi^2 z^4} \right)
\;\; = \;\; -i \delta^{(4)} (z) ,
\eeq
one can easily show that the twist--2 part of the
current product satisfies
\beq
\frac{\partial}{\partial z_\mu} \Pi_{\{\mu\nu\}}^{(0)}
\;\; = \;\; 0 , 
\hspace{5em}
\frac{\partial}{\partial z_\mu} \Pi_{[\mu\nu ]}^{(0)}
\;\; = \;\; 0 ,
\label{transversality_twist_2}
\eeq
where 
\beq
\left. \begin{array}{c} \Pi_{\{\mu\nu\}}^{(0)}  \\[1.5ex]
\Pi_{[\mu\nu ]}^{(0)} \end{array}  \right\} 
\;\; \equiv \;\; 
\frac{1}{2} \left[ \Pi_{\mu\nu}^{(0)} \pm \Pi_{\nu\mu}^{(0)} \right]
\eeq
denote the Lorentz--tensor symmetric and antisymmetric parts.
These properties play an important role in ensuring transversality of the 
light--cone expansion, as will be shown in the following subsection.
\subsection{Transversality and twist--3 operators}
\label{subsec_transversality}
{\it Transversality of the twist--2 contribution.}
We now turn to 
the question how to maintain transversality (electromagnetic
gauge invariance) in the light--cone expansion of the current product.
For the operator product, Eq.(\ref{Pi}), regarded as a function of the
two points, $x$ and $y$, current conservation implies that
\be
\frac{\partial}{\partial x_\mu} \Pi_{\mu\nu} (x, y) \;\; = \;\; 0,
\hspace{2cm}
\frac{\partial}{\partial y_\nu } \Pi_{\mu\nu} (x, y)
\;\; = \;\; 0 .
\label{transversality_x_y}
\ee
In terms of the center and relative coordinates, 
Eq.(\ref{COM_relative_coordinates}), these conditions take the 
form
\be
\frac{\partial}{\partial z_\mu} \Pi_{\{\mu\nu\}} (z|X) 
&=& \frac{1}{2} \frac{\partial}{\partial X_\mu} \Pi_{[\mu\nu ]} (z|X) ,
\label{transversality_sym}
\\
\frac{\partial}{\partial z_\mu} \Pi_{[\mu\nu ]} (z|X)
&=& \frac{1}{2} \frac{\partial}{\partial X_\mu} 
\Pi_{\{\mu\nu\}} (z|X) .
\label{transversality_antisym}
\ee
One sees that the transversality conditions relate the Lorentz--tensor 
symmetric and antisymmetric parts of the operator product; 
{\it cf.}\ Eq.(\ref{trans}) for the Compton amplitude.
\par
In DIS the twist--2 part of the light--cone expansion,
Eq.(\ref{handbag_twist_2}), gives the leading contribution to the structure
functions at large $q^2$. In this case one is dealing with forward matrix
elements of the twist--2 operators, so that all total derivatives of
operators ({\it i.e.}, derivatives $\partial / \partial X$) have zero
matrix elements. The two conditions, Eqs.(\ref{transversality_sym}) and
(\ref{transversality_antisym}), decouple, and the properties
Eq.(\ref{transversality_twist_2}) are sufficient to guarantee transversality
of both the symmetric and antisymmetric parts separately.  However, one
immediately sees that in the case of non-forward matrix elements, as 
{\it e.g.}\ in DVCS, the twist--2 part alone is not transverse. This is
because in the non-forward case total derivatives of operators generally
have non-zero matrix element, and one can easily verify that in general
\beq
\frac{\partial}{\partial X_\mu} \Pi_{\{\mu\nu\}}^{(0)}
\;\; \neq \;\; 0 ,
\hspace{5em}
\frac{\partial}{\partial X_\mu} \Pi_{[\mu\nu ]}^{(0)}
\;\; \neq \;\; 0 .
\eeq
As will be shown in Section~\ref{sec_amplitude}, hadronic matrix elements 
of these total derivatives for DVCS kinematics are 
generally of order unity, {\it i.e.}, not suppressed by factors 
$m^2/q^2$ or $t/q^2$. Thus, the violation of transversality by the
twist--2 contribution alone cannot be neglected at the accuracy we are
aiming for. Since the total operator product obviously is transverse, 
our conclusion can only be that in order to maintain transversality
to the required accuracy we have to include operators
of twist $> 2$ in the light--cone expansion.
\par
{\it Including higher--twist operators.}
It turns out that the higher--twist operators needed to maintain
transversality are contained already in the simple ``handbag'' 
term of the string operator expansion, Eq.(\ref{handbag_string}); 
the subleading terms in $1/z^2$ in the 
string operator expansion are not needed for this. The necessary
operators are total derivatives of twist--2 operators, 
which are algebraically of twist 3. They naturally appear when taking 
into account the ``rest'' of the original
string operator, which was dropped in taking the twist--2 part, 
and making use of the QCD equations of motion.
\par
In order to account for the twist $>2$ parts of the string 
operators in Eq.(\ref{handbag_string}) we first note that the differences
between the symmetrized operators, Eq.(\ref{string_sym}), and the full 
string operators can be expressed in the form 
(for brevity we put $X = 0$ in the following) \cite{Balitsky:1989bk}
\be
\lefteqn{
\bar\psi (-z/2) \gamma_\sigma \psi (z/2)
\; - \; \left[ \bar\psi (-z/2) \gamma_\sigma \psi (z/2) \right]^{\rm sym} }
&& \nonumber \\
&=& \int_0^1 dt\; z_\alpha 
\left[ \frac{\partial}{\partial z_\alpha} 
\bar\psi (-tz/2) \gamma_\sigma \psi (tz/2)
\; - \; (\sigma \leftrightarrow \alpha) \right] ,
\label{rest_general}
\ee 
and similarly for the operator with Dirac matrix 
$\gamma_\sigma \gamma_5$. To show this one uses that for any 
function, $f(tz)$,
\beq
z_\alpha \frac{\partial}{\partial z_\alpha} f(tz) 
\;\; = \;\; t \frac{\partial}{\partial t} f(tz) . 
\label{z_dz_identity}
\eeq
The derivatives of the string operator in Eq.(\ref{rest_general}) can
be computed with the help of Eq.(\ref{string_derivative_z}). Of the
terms with covariant derivatives of the fermion fields at the end points
one vanishes by virtue of the QCD equations of motion,
\beq
\ldots \gamma_\alpha \nablaright_\alpha \psi (x) \;\; = \;\; 0 , 
\hspace{5em}
\psi (x) \nablaleft_\alpha \gamma_\alpha  \ldots \;\; = \;\; 0 , 
\eeq
the other, with the help of Eq.(\ref{string_derivative_X}), can be traded 
for a derivative with respect to the center coordinate, plus additional 
quark--gluon string operators. In this way one 
arrives at \cite{Balitsky:1989bk}
\be
\lefteqn{
\bar\psi (-z/2) \gamma_\sigma \psi (z/2)
\; - \; \left[ \bar\psi (-z/2) \gamma_\sigma \psi (z/2) \right]^{\rm sym} }
&& \nonumber \\
&=& \frac{i}{2} \epsilon_{\sigma\alpha\beta\gamma} z_\alpha
\frac{\partial}{\partial X_\beta} \int_0^1 dt \, t \;
\bar\psi (-tz/2) \gamma_\gamma \gamma_5  \psi (tz/2)
\nonumber \\
&+& \mbox{string operators of type 
$\bar\psi \ldots F \ldots \psi$} .
\label{rest_total_derivative}
\ee
A corresponding relation holds for the operator with Dirac matrix 
$\gamma_\sigma\gamma_5$ on the L.H.S.\ and the operator 
with $\gamma_\gamma$ on the R.H.S.:
\be
\lefteqn{
\bar\psi (-z/2) \gamma_\sigma \gamma_5 \psi (z/2)
\; - \; \left[ \bar\psi (-z/2) \gamma_\sigma \gamma_5 
\psi (z/2) \right]^{\rm sym} }
&& \nonumber \\
&=& \frac{i}{2} \epsilon_{\sigma\alpha\beta\gamma} z_\alpha
\frac{\partial}{\partial X_\beta} \int_0^1 dt \, t \;
\bar\psi (-tz/2) \gamma_\gamma \psi (tz/2)
\nonumber \\
&+& \mbox{string operators of type 
$\bar\psi \ldots F \ldots \psi$} .
\label{rest_5_total_derivative}
\ee
The precise form of the quark--gluon operators appearing here need not
concern us; it turns out that these parts do not play a role in restoring
transversality of the light--cone expansion.  Note that
Eqs.(\ref{rest_total_derivative}) and (\ref{rest_5_total_derivative})
relate the vector and axial vector string operators, which is possible
with the help of the totally antisymmetric $\epsilon$--tensor. We emphasize
that the operators appearing under the total derivative on the R.H.S. of
Eqs.(\ref{rest_total_derivative}) and (\ref{rest_5_total_derivative}) are
still full string operators, which have as yet no definite twist. In
particular, they may again be decomposed into a symmetric ({\it i.e.},
twist--2) part and total derivatives, and so on. In fact,
Eqs.(\ref{rest_total_derivative}) and (\ref{rest_5_total_derivative}) can
be regarded as recurrence relations, repeated use of which allows to
express the original string operator as the sum of its symmetric part and
an infinite series of total derivatives of symmetric operators of arbitrary
order (and quark--gluon operators, which we do not consider explicitly). We
shall show below that these recurrence relations can be solved, and that
the series of total derivative operators can be summed up in
closed form.
\par
{\it First--order correction in total derivatives.} 
In order to understand how the total derivatives appearing in 
Eqs.(\ref{rest_total_derivative}) and 
(\ref{rest_5_total_derivative}) come into play in restoring 
the transversality of the light--cone expansion, it is 
instructive to first take a look at the first--order correction in 
total derivatives to the twist--2 part of the operator product. 
To first order in $\partial / \partial X$ we have (we do not write
the contributions from quark--gluon string operators any longer)
\be
\bar\psi (-z/2) \gamma_\sigma \psi (z/2)
&=& \left[ \bar\psi (-z/2) \gamma_\sigma \psi (z/2) \right]^{\rm sym}
\nonumber \\
&+& \frac{i}{2} \epsilon_{\sigma\alpha\beta\gamma} z_\alpha
\frac{\partial}{\partial X_\beta} \int_0^1 dt \, t \;
\left[ \bar\psi (-tz/2) \gamma_\gamma \gamma_5  \psi (tz/2) \right]^{\rm sym}
\nonumber \\
&+& \mbox{terms of order} \;\; \frac{\partial}{\partial X} 
\frac{\partial}{\partial X} ,
\label{string_first_order}
\ee
and similarly for the operators with 
$\gamma_\sigma \rightarrow \gamma_\sigma\gamma_5, 
\gamma_\gamma\gamma_5 \rightarrow \gamma_\gamma$. When substituted
in Eq.(\ref{handbag_string}), the first--order corrections to the
twist--2 string operators produce a contribution to the operator 
product which we denote by $\Pi_{\mu\nu}^{(1)}$:
\be
\Pi^{(1)}_{\mu\nu} &=& \frac{z_\rho}{\pi^2 z^4}
\left\{ i (z_\mu g_{\sigma\nu} + z_\nu g_{\sigma\mu})
\varepsilon_{\sigma\rho\beta\gamma} 
\frac{\partial}{\partial X_\beta} 
\int_0^1 dt\, t\; 
\left[ \bar\psi (-tz/2) \gamma_\gamma \gamma_5  
\psi (tz/2) \; + \; (z \rightarrow -z) \right]^{\rm sym}
\right.
\nonumber \\
&& \phantom{\frac{z_\rho}{4 \pi^2 z^4}} 
\left. + \; \varepsilon_{\sigma\rho\beta\gamma}
\varepsilon_{\mu\nu\sigma\delta} z_\delta 
\frac{\partial}{\partial X_\beta}
\int_0^1 dt\, t\;
\left[ \bar\psi (-tz/2) \gamma_\gamma
\psi (tz/2) \; - \; (z \rightarrow -z) \right]^{\rm sym}
\right\} . \;\;\;\;\;\;\;
\label{Pi_1}
\ee
Let us see what the inclusion of this first--order correction in 
$\partial / \partial X$ to the current product implies for
the transversality conditions, Eqs.(\ref{transversality_sym})
and (\ref{transversality_antisym}). By direct calculation one can easily 
show that
\be
\frac{\partial}{\partial z_\mu} \Pi_{\{\mu\nu\}}^{(1)}
&=& \frac{1}{2} \frac{\partial}{\partial X_\mu} \Pi_{[\mu\nu ]}^{(0)}
\label{transversality_sym_first_order}
\\
\frac{\partial}{\partial z_\mu} \Pi_{[\mu\nu ]}^{(1)}
&=& \frac{1}{2} \frac{\partial}{\partial X_\mu} 
\Pi_{\{\mu\nu\}}^{(0)} .
\label{transversality_antisym_first_order}
\ee
The equalities here are understood up to terms involving contracted total 
derivatives acting on the symmetric string operator,
\beq
\frac{\partial}{\partial X_\mu} \frac{\partial}{\partial X_\mu} 
\left[ \bar\psi (-z/2) \gamma_\sigma \psi (z/2) \right]^{\rm sym} ,
\label{total_derivative_squared}
\eeq
or ``total divergences'' of the symmetric string operator, 
\beq
\frac{\partial}{\partial X_\sigma} 
\left[ \bar\psi (-z/2) \gamma_\sigma \psi (z/2) \right]^{\rm sym} .
\label{total_divergence}
\eeq
The hadronic matrix elements of derivatives of type
Eq.(\ref{total_derivative_squared}) will be proportional to the square of
the momentum transfer, $t$, so their contribution to the non-forward
Compton amplitude will be suppressed by $t/q^2$. Similarly, derivatives of
type Eq.(\ref{total_divergence}) will have matrix elements proportional to
either $m^2$ or $t$, and thus again lead to power--suppressed contributions
to the Compton amplitude.\footnote{For $m^2 = 0$ and $t = 0$ the matrix
element of the twist--2 vector string operator has only contributions
proportional to the momenta, $p_\sigma$ and $r_\sigma$. Terms proportional
to $z_\sigma$, which originate from $z^2$--terms in the contracted
operator, Eq.(\ref{string_sym}), come with factors of $m^2$ or $t$.}
\par
To verify Eq.(\ref{transversality_sym_first_order}) one first shows that
\be
\frac{\partial}{\partial z_\mu} \Pi_{\{\mu\nu\}}^{(1)}
&=& - \frac{z_\rho}{\pi^2 z^4} \varepsilon_{\nu\rho\beta\gamma} 
\frac{\partial}{\partial X_\beta} 
\left[ 2 + \left( z \frac{\partial}{\partial z} \right) \right]
\nonumber \\
&& \times \int_0^1 dt\, t \, \left[ \bar\psi (-tz/2) \gamma_\gamma \gamma_5  
\psi (tz/2) \; + \; (z \rightarrow -z) \right]^{\rm sym} ,
\ee
and notes that the derivative $z_\alpha (\partial / \partial z_\alpha )$ 
acting on the string operator of argument $tz$ can be replaced by 
$t (\partial /\partial t)$, {\it cf.}\ Eq.(\ref{z_dz_identity}).
The R.H.S.\ of Eq.(\ref{transversality_sym_first_order}) is then
obtained after integration by parts in the parameter, $t$.
To verify Eq.(\ref{transversality_antisym_first_order}) one
has to make use of the fact the fact that, because of 
Eq.(\ref{string_sym}), the symmetric 
part of the string operator satisfies 
\beq
\frac{\partial}{\partial z_\mu} 
\left[ \bar\psi (-tz/2) \gamma_\gamma \gamma_5  
\psi (tz/2) \right]^{\rm sym}
\;\; = \;\;
\frac{\partial}{\partial z_\gamma} 
\left[ \bar\psi (-tz/2) \gamma_\mu \gamma_5  
\psi (tz/2) \right]^{\rm sym} ,
\eeq
and drop contributions of the type of
Eqs.(\ref{total_derivative_squared}) and (\ref{total_divergence}).
\par
{\it Restoring transversality order--by--order in total derivatives.} 
Is the inclusion of the first--order correction in total derivatives,
Eq.(\ref{Pi_1}), in addition to the twist--2 contribution sufficient to 
restore transversality? Substituting
\beq
\Pi_{\mu\nu}^{(0)} \; + \; \Pi_{\mu\nu}^{(1)}
\eeq
into the transversality conditions, Eqs.(\ref{transversality_sym}) 
and (\ref{transversality_antisym}), and making use of the 
results Eq.(\ref{transversality_twist_2}) for the twist--2 part and 
Eqs.(\ref{transversality_sym_first_order}), 
(\ref{transversality_antisym_first_order}) for the first--order correction,
we find
\be
\frac{\partial}{\partial z_\mu} \left[ 
\Pi_{\{\mu\nu\}}^{(0)} + \Pi_{\{\mu\nu\}}^{(1)} \right]
\; - \; \frac{1}{2} \frac{\partial}{\partial X_\mu} 
\left[ \Pi_{[\mu\nu ]}^{(0)} + \Pi_{[\mu\nu ]}^{(1)} \right] 
&=& - \frac{1}{2} \frac{\partial}{\partial X_\mu} \Pi_{[\mu\nu ]}^{(1)} ,
\\
\frac{\partial}{\partial z_\mu} \left[ 
\Pi_{[\mu\nu ]}^{(0)} + \Pi_{[\mu\nu ]}^{(1)} \right]
\; - \; \frac{1}{2} \frac{\partial}{\partial X_\mu} 
\left[ \Pi_{\{\mu\nu\}}^{(0)} + \Pi_{\{\mu\nu\}}^{(1)} \right] 
&=& - \frac{1}{2} \frac{\partial}{\partial X_\mu} \Pi_{\{\mu\nu\}}^{(1)} .
\ee
Although the R.H.S.\ is of second order in total derivatives, it is {\it
not} suppressed in the approximation we are considering, {\it i.e.}, it is
not proportional to derivatives of type
Eqs.(\ref{total_derivative_squared}) or (\ref{total_divergence}), which
give power--suppressed contributions to the Compton amplitude.  In fact,
one can easily see that the ``additional'' total derivative in the
operators on the R.H.S.\ appears in the combination 
$z_\mu (\partial / \partial X_\mu )$, which under the hadronic matrix
element becomes $-i(rz)$ and is not suppressed in DVCS kinematics. Thus,
compared to the twist--2 part alone we have merely shifted the violation of
transversality to terms one order higher in 
$z_\mu (\partial / \partial X_\mu ) \, = \, -i(rz)$. We conclude that, in
order to obtain a transverse result, we have to include the total
derivative operators to {\it all} orders:
\beq
\Pi_{\mu\nu}^{\rm transverse} \;\; = \;\; 
\Pi_{\mu\nu}^{(0)} \; + \; \Pi_{\mu\nu}^{(1)} \; + \; \Pi_{\mu\nu}^{(2)}
\; + \ldots ,
\label{Pi_transverse_def}
\eeq
where the superscript $(n)$ denotes the order in total
derivatives.\footnote{We stress again that the ``order'' in total
derivatives here does not imply suppression of the contributions in
$t/q^2$.}  These contributions are generated by successive use of the
recurrence relations, Eqs.(\ref{rest_total_derivative}) and
(\ref{rest_5_total_derivative}), which allow to express the ``rest'' of a
symmetrized string operator in terms of total derivatives and quark--gluon
string operators. (The latter turn out not to play a role in restoring
transversality, so we shall drop them.) The transversality conditions
connect the term of order $n$ in total derivatives with that of order 
$n + 1$; {\it i.e.}, instead of Eq.(\ref{transversality_sym_first_order})
and (\ref{transversality_antisym_first_order}) one now has
\be
\frac{\partial}{\partial z_\mu} \Pi_{\{\mu\nu\}}^{(n + 1)}
&=& \frac{1}{2} \frac{\partial}{\partial X_\mu} \Pi_{[\mu\nu ]}^{(n)} ,
\\
\frac{\partial}{\partial z_\mu} \Pi_{[\mu\nu ]}^{(n + 1)}
&=& \frac{1}{2} \frac{\partial}{\partial X_\mu} 
\Pi_{\{\mu\nu\}}^{(n)} .
\ee
\par
{\it All--order summation of total derivatives.}
Iterating Eqs.(\ref{rest_total_derivative}) and 
(\ref{rest_5_total_derivative}), we express the string operators
appearing in Eq.(\ref{handbag_string}) as the sum of its
symmetric part, Eq.(\ref{string_sym}), and an infinite
series of total derivatives of symmetric string operators.
This series can be summed up in closed form. The details of the calculation
are presented in Appendix \ref{app_deconstructing}. The result for the 
``fully deconstructed'' string operators takes the form:
\be
\lefteqn{
\bar\psi (-z/2) \gamma_\sigma \psi (z/2) }
&& \nonumber \\
&=& \left[ \bar\psi (-z/2) \gamma_\sigma \psi (z/2) \right]^{\rm sym}
\nonumber \\
&+& \frac{i}{2} \left[ \frac{\partial}{\partial X_\sigma} z_\gamma 
- \left( z \frac{\partial}{\partial X} \right) g_{\sigma\gamma} \right]
\; \int_0^1 du \, u \; \sin \left[ \frac{i \bar u}{2} 
\left( z \frac{\partial}{\partial X} \right) \right]
\left[ \bar\psi (-uz/2) \gamma_\gamma  \psi (uz/2) \right]^{\rm sym}
\nonumber \\
&+& \frac{i}{2} \epsilon_{\sigma\alpha\beta\gamma} z_\alpha 
\frac{\partial}{\partial X_\beta}
\int_0^1 du \, u \; \cos \left[ \frac{i \bar u}{2} 
\left( z \frac{\partial}{\partial X} \right) \right] 
\left[ \bar\psi (-uz/2) \gamma_\gamma \gamma_5 \psi (uz/2) 
\right]^{\rm sym} \;\;\;\;
\nonumber \\[1ex]
&+& \mbox{contracted operators of the type 
(\ref{total_derivative_squared}) or (\ref{total_divergence})}
\nonumber \\[1ex]
&+& \mbox{quark--gluon operators} ,
\label{string_deconstructed}
\ee
and similarly for the operators with 
$\gamma_\sigma \rightarrow \gamma_\sigma\gamma_5, 
\gamma_\kappa\gamma_5 \rightarrow \gamma_\kappa$. [Note that when 
expanding to first order in $\partial / \partial X$ we recover 
Eq.(\ref{string_first_order}).] We shall refer to the part
of the string operator involving only symmetrized operators and
their total derivatives as
\beq
\left[ \bar\psi (-z/2) \gamma_\sigma \psi (z/2) 
\right]^{\mbox{\scriptsize sym + total der.}} .
\eeq
In Eq.(\ref{string_deconstructed}) we can substitute on the R.H.S.\ the
explicit representation for the symmetric string operators, 
Eq.(\ref{string_sym}). Performing an integration by parts over $u$
in the second term of Eq.(\ref{string_deconstructed}), and using
the identity (here $v \equiv u t$)
\beq
\int_0^1\! du \int_0^1\! dt \, f(u) \, g(ut) \;\; = \;\; 
\int_0^1\! du \int_0^u\! 
dv \, f(u) \, g(v)
\;\; = \;\; \int_0^1\! dv \, g(v) \left[ \int_v^1\! du \, f(u)\right]
\eeq
to simplify the third term, we obtain
\be
\lefteqn{
\left[ \bar\psi (-z/2) \gamma_\sigma \psi (z/2) 
\right]^{\mbox{\scriptsize sym + total der.}} }
&& \nonumber \\
&=& \int_0^1 dv \; \left\{ \cos \left[ \frac{i \bar v}{2} 
\left( z \frac{\partial}{\partial X} \right) \right]
\frac{\partial}{\partial z_\sigma}
\; + \; \frac{i v}{2}
\sin \left[ 
\frac{i \bar v}{2} \left( z \frac{\partial}{\partial X} \right) \right]
\frac{\partial}{\partial X_\sigma}
\right\} \bar\psi (-vz/2) \hat z \psi (vz/2) 
\nonumber \\
&+& \frac{i}{2} \epsilon_{\sigma\alpha\beta\gamma} z_\alpha 
\frac{\partial}{\partial X_\beta} \frac{\partial}{\partial z_\gamma}
\int_0^1 dv \int_v^1 du \;
\cos \left[ \frac{i \bar u}{2} 
\left( z \frac{\partial}{\partial X} \right) \right] 
\; \bar\psi (-vz/2) \hat z \gamma_5 \psi (vz/2) . 
\label{string_deconstructed_scalar}
\ee
The parameter integral in the second term can be carried out and gives
\beq
\int_v^1 du 
\cos \left[ \frac{i \bar u}{2} \; 
\left( z \frac{\partial}{\partial X} \right) \right] 
\;\; = \;\; \frac{\displaystyle 
\sin \left[ \frac{i \bar v}{2} \left( 
z \frac{\partial}{\partial X} \right) \right]}
{\displaystyle \frac{i}{2} \left( z \frac{\partial}{\partial X}\right)} .
\eeq
An analogous formula applies to the operators with
$\gamma_\sigma \rightarrow \gamma_\sigma\gamma_5$; on the R.H.S.\ one
should replace $\hat z \rightarrow \hat z \gamma_5 , \; \hat z \gamma_5 
\rightarrow \hat z$. Note that to zeroth order in total derivatives,
the R.H.S.\ of Eq.(\ref{string_deconstructed_scalar}) just reduces to 
the usual symmetrized operator, Eq.(\ref{string_sym}).
\par
{\it Finite translation of symmetric string operators.}
The sine and cosine functions involving the total derivative, appearing in 
Eq.(\ref{string_deconstructed}), are nothing but the odd and even part
of the finite translation operator
\beq
\exp \left[ \pm \frac{\bar v}{2} 
\left( z\frac{\partial}{\partial X}\right) \right] ,
\eeq
which affects a translation of the center of the string operators by a 
distance $\pm \bar v z/2$. Taking this into account, we may alternatively
write the result for the string operator containing total derivatives
of symmetrized operators, Eq.(\ref{string_deconstructed_scalar}), 
in the form
\be
\lefteqn{
\left[ \bar\psi (-z/2) \gamma_\sigma \psi (z/2) 
\right]^{\mbox{\scriptsize sym + total der.}} }
&& \nonumber \\
&=& \frac{1}{2} \int_0^1 dv \left\{
\left( 
\frac{\partial}{\partial z_\sigma} + \frac{v}{2} 
\frac{\partial}{\partial X_\sigma}
\right) \bar\psi (-z/2) \, \hat z \,\, \psi [(2v - 1) z/2] \right.
\nonumber \\
&& \left. \phantom{\frac{1}{2} \int_0^1 dv}\!\! + \left( 
\frac{\partial}{\partial z_\sigma} - \frac{v}{2} 
\frac{\partial}{\partial X_\sigma}
\right) \bar\psi [-(2v - 1) z/2] \,\, \hat z \, \psi (z/2) \right\}
\nonumber \\
&+& \frac{i}{4} \epsilon_{\sigma\alpha\beta\gamma} z_\alpha 
\frac{\partial}{\partial X_\beta} \frac{\partial}{\partial z_\gamma}
\; \int_0^1 dv \int_v^1 du 
\left\{ 
\bar\psi [-(v + \bar u) z/2] \,\, \hat z \, \gamma_5 \psi [(v - \bar u )z/2]
\phantom{\int_0^1 dv} \right.
\nonumber \\
&& \left. \phantom{\int_0^1 dv \frac{i}{4} 
\epsilon_{\sigma\alpha\beta\gamma} z_\alpha 
\frac{\partial}{\partial X_\beta} \frac{\partial}{\partial z_\gamma}
\; \int_v^1 du}
+ \bar\psi [-(v - \bar u) z/2] \,\, \hat z \,
\gamma_5 \psi [(v + \bar u )z/2]
\right\} .
\label{string_deconstructed_translation}
\ee
We see that taking into account total derivatives of symmetrized operators 
to all orders in the decomposition of the string operator amounts to
certain finite translations of the symmetrized operators,
along the direction defined by the separation of the fields in the 
original operator, $z$. It is interesting to note that the endpoints of 
the symmetrized string operators under the parameter integrals lie 
inside the interval $[-z/2, z/2]$ defined by the original operator;
this can be regarded as a consequence of locality.\footnote{We are grateful
to V.M.~Braun for bringing this point to our attention.}
\par
{\it Transverse extension of leading light--cone singularity.}
Substituting the closed--form expression for the decomposition of the string 
operator in symmetric operators and total derivatives thereof
into Eq.(\ref{handbag_string}) we can now obtain the expression
for the transverse extension of the twist--2 part of the light--cone 
expansion, Eq.(\ref{Pi_transverse_def}):
\be
\Pi_{\mu\nu}^{\rm transverse}
&=& \frac{2 z_\rho}{\pi^2 z^4} \left\{ \phantom{-i}
s_{\mu\rho\nu\sigma} 
\left[\bar\psi (-z/2) \gamma_\sigma \psi (z/2) 
\phantom{\gamma_5} - (z \rightarrow -z)
\right]^{\mbox{\scriptsize sym + total der.}}  \right.
\nonumber \\
&& \phantom{\frac{z_\rho}{2\pi^2 z^4}} \left. 
-i \epsilon_{\mu\rho\nu\sigma} 
\left[\bar\psi (-z/2) \gamma_\sigma \gamma_5 
\psi (z/2) + (z \rightarrow -z)
\right]^{\mbox{\scriptsize sym + total der.}} 
\right\} .
\label{handbag_transverse}
\ee
One can verify that this expression satisfies the transversality
conditions, Eqs.(\ref{transversality_sym}) and
(\ref{transversality_antisym}), up to terms involving operators of the type
Eqs.(\ref{total_derivative_squared}) or (\ref{total_divergence}), which
give power--suppressed contributions to the Compton amplitude.  [To see
this, one needs to make use of the identity Eq.(\ref{z_dz_identity}) and
integrate by parts over the parameter $v$.] We do not bother to make the
string operators traceless here. As explained above, $z^2$--terms in the
operators need to be included explicitly only when keeping corrections of
order $t/q^2$ and $m^2/q^2$ to the DVCS amplitude. Note that in this case
one would need to include also operators of the form
Eqs.(\ref{transversality_sym}) and (\ref{transversality_antisym}) in the
decomposition of the string operators, Eq.(\ref{string_deconstructed}).
\par
We can substitute in Eq.(\ref{handbag_transverse}) the explicit
expressions for the vector string operators in terms of the 
contracted string operators, Eq.(\ref{string_deconstructed_scalar}), 
and express our result for the transverse generalization of 
the leading term light--cone expansion directly in terms of the 
contracted operators. Grouping together contributions from the 
operators with Dirac structures $\hat z$ and $\hat z \gamma_5$ we obtain
\be
\Pi_{\mu\nu}^{\rm transverse}
&=& \frac{2}{\pi^2 z^4} \int_0^1 dv \; \left\{ 
\cos \left[ \frac{i \bar v}{2} 
\left( z \frac{\partial}{\partial X} \right) \right]
s_{\mu\rho\nu\sigma} z_\rho \frac{\partial}{\partial z_\sigma}
\; + \; \frac{iv}{2}
\sin \left[ \frac{i \bar v}{2} \left( z \frac{\partial}{\partial X} \right) 
\right]
s_{\mu\rho\nu\sigma} z_\rho \frac{\partial}{\partial X_\sigma}
\right.
\nonumber \\
&& \left. \phantom{\frac{2}{\pi^2 z^4} \int_0^1 dv} \;
+ \, \frac{1}{2}  
\; \int_v^1 du 
\cos \left[ \frac{i \bar u}{2} \left( z \frac{\partial}{\partial X} 
\right) \right]
\epsilon_{\mu\rho\nu\sigma} 
\epsilon_{\sigma\alpha\beta\gamma} z_\rho z_\alpha 
\frac{\partial}{\partial X_\beta}
\frac{\partial}{\partial z_\gamma} \right\}
\nonumber \\
&& \phantom{\frac{2}{\pi^2 z^4} \int_0^1 dv}
\times \left[ \bar\psi (-vz/2) \hat z \psi (vz/2) 
\; + \; (z \rightarrow -z) \right] 
\nonumber 
\\[1.5ex]
&+& \frac{2}{\pi^2 z^4} \int_0^1 dv \; \left\{ 
-i \cos \left[ \frac{i \bar v}{2} \left( 
z \frac{\partial}{\partial X} \right) \right]
\epsilon_{\mu\rho\nu\sigma} z_\rho \frac{\partial}{\partial z_\sigma}
\; + \; \frac{v}{2}
\sin \left[ \frac{i \bar v}{2} \left( z 
\frac{\partial}{\partial X} \right) \right]
\epsilon_{\mu\rho\nu\sigma} z_\rho \frac{\partial}{\partial X_\sigma}
\right.
\nonumber \\
&& \left. \phantom{\frac{2}{\pi^2 z^4} \int_0^1 dv} \;
+ \frac{i}{2} \int_v^1 du 
\cos \left[ \frac{i \bar u}{2} \left( z \frac{\partial}{\partial X} \right) 
\right] (\epsilon_{\mu\alpha\beta\gamma} z_\nu 
+ \epsilon_{\nu\alpha\beta\gamma} z_\mu) z_\alpha   
\frac{\partial}{\partial X_\beta} 
\frac{\partial}{\partial z_\gamma} \right\}
\nonumber \\
&& \phantom{\frac{2}{\pi^2 z^4} \int_0^1 dv}
\times
\left[ \bar\psi (-vz/2) \hat z \gamma_5 \psi (vz/2) 
\; - \; (z \rightarrow -z) \right] .
\label{handbag_transverse_alt}
\ee
This form is useful because the two parts involving the operators 
$\bar\psi \hat z \psi$ and $\bar\psi \hat z \gamma_5 \psi$ turn
out to be {\it individually} transverse to the accuracy stated above.
(Note that both these terms contain a symmetric and an antisymmetric tensor part.) 
This is necessarily so: There is in general no dynamical relation between the 
matrix elements of the operator $\bar\psi \hat z \psi$ and those
of the operator $\bar\psi \hat z \gamma_5 \psi$ (in the forward limit they 
are related, respectively, to the unpolarized and polarized 
quark distributions in the hadron), so their contributions to the
light--cone expansion must be individually transverse.
\subsection{Quark matrix elements: Recovering the handbag diagram}
\label{subsec_quark}
Before turning to the hadronic DVCS amplitude, it is instructive to compute 
the matrix element of the leading term in the light--cone expansion, 
Eq.(\ref{handbag_transverse}), between quark states. This simple exercise
serves two purposes. First, we can verify that the Compton amplitude
for a quark target, obtained from Eq.(\ref{handbag_transverse}), is --- up
to terms of order $t/q^2$ --- equivalent
to the result of the simple ``handbag'' graph in a theory of free quarks.
This may not come as a surprise, as the operator result,
Eq.(\ref{handbag_transverse}), was obtained by way of a reduction
of the simplest two--point string operator, whose graphical analogue in 
the free theory would just be the Compton amplitude for the free quark. 
Still, it is worthwhile to check explicitly that the operators of 
twist $> 3$, which were dropped in the process of reduction of the 
string operator, are not neeeded to reproduce the ``handbag'' result. 
Second, when computing the quark Compton amplitude
from Eq.(\ref{handbag_transverse}) we shall see that characteristic
singularities appear in terms with denominators quartic in momenta, 
which are a consequence of the finite translations of the string
operators in the gauge--invariant expression, {\it cf.}\
Eqs.(\ref{string_deconstructed_scalar}) and 
(\ref{string_deconstructed_translation}). In the free quark case these 
singularities cancel when adding the contributions from the two twist--2
operators, $\bar\psi \hat z \psi$ and $\bar\psi \hat z \gamma_5 \psi$.
Similar singularities we shall also encounter in the DVCS amplitude for 
a hadronic target; however, in this case they do not cancel completely, 
leaving certain divergences in the tensor amplitude for transverse photon 
polarization (see Section~\ref{sec_amplitude}).
\par
{\it Quark matrix elements of symmetrized string operators.}
Let us compute the matrix element of Eq.(\ref{handbag_transverse}) 
between free massless quark states. Their 
wave functions are plane waves of momenta $p\pm r/2$, multiplied by Dirac 
spinors satisfying
\beq
\left( \hat p + \frac{\hat r}{2} \right) \,\, U \;\; = \;\; 0 ,
\hspace{5em}
\bar U \,\, \left( \hat p - \frac{\hat r}{2} \right) \;\; = \;\; 0 .
\label{Dirac_equation_quark}
\eeq
The quark matrix elements of the totally symmetric (twist--2)
light--ray operators (with center coordinate $X = 0$) 
are
\beq
\langle p - r/2| \bar\psi (-z/2) 
\left\{ \begin{array}{c} \hat z \\[1.5ex] 
\hat z \gamma_5 \end{array}  \right\} 
\psi (z/2) | p + r/2 \rangle
\;\; = \;\; 
\bar U \left\{ \begin{array}{c} \hat z \\[1.5ex] 
\hat z \gamma_5 \end{array}  \right\} U \,
e^{-i (pz)} .
\label{me_quark_general}
\eeq
Only the average momentum, $p$, enters in 
the phase factor containing the separation of the fields, $z$, in accordance 
with translational invariance. From these one can easily construct
the matrix elements of the vector and axial vector string operators
including symmetrized operators and their total derivatives, 
Eq.(\ref{string_deconstructed_scalar}). We find:
\be
\lefteqn{
\langle p - r/2| \left[ \bar\psi (-z/2) \gamma_\sigma \psi (z/2) 
\right]^{\mbox{\scriptsize sym + total der.}} 
| p + r/2 \rangle}
\nonumber \\
&=& e^{-i (pz)} \; \bar U \gamma_\sigma U 
\nonumber \\
&+& \int_0^1 dv \, e^{-i v (pz)} \; 
\left\{ \;\; i v \cos [ \bar v (rz)/2 ] \; 
\bar U \left[ (pz) \gamma_\sigma - p_\sigma \hat{z} \right] U
\phantom{\int_0^1} \right.
\nonumber \\
&& \phantom{\int_0^1 dv \, e^{-i v (pz)} \;}
- \, \frac{v}{2} \sin [ \bar v (rz)/2 ] \;
\bar U \left[ (rz) \gamma_\sigma - r_\sigma \hat{z} \right] U
\nonumber \\
&& \left. \phantom{\int_0^1 dv \, e^{-i v (pz)} \;}
+ \; \frac{1}{2} \epsilon_{\sigma\alpha\beta\gamma} z_\alpha r_\beta
\int_v^1 du\, \cos [ \bar u (rz)/2 ] \; 
\bar U \left( \gamma_\gamma - i v p_\gamma \hat{z} \right) \gamma_5 U
\right\} .
\label{vector_quark}
\ee
Here the first three terms originate from the contracted operator
$\bar\psi \hat z \psi$, the fourth from 
$\bar\psi \hat z \gamma_5 \psi$. We have performed an integration by 
parts in the integral over the parameter $v$ in the term proportional 
to $\bar U \gamma_\sigma U$, in order to isolate the first term, which
coincides with the free quark matrix element of the vector string 
operator, $\bar\psi (-z/2) \gamma_\sigma \psi (z/2)$. Note that
all terms but this one vanish in the limit of zero momentum transfer,
$r \rightarrow 0$. In order to simplify the second and third term we 
make use of the following identities for quark bilinears, which follow from
the Dirac equation for the quark spinors, Eq.(\ref{Dirac_equation_quark}), 
and the identities for products of three gamma matrices:
\be
\bar U \left[ (pz) \gamma_\sigma - p_\sigma \hat{z} \right] U &=& 
\frac{i}{2} \epsilon_{\sigma\alpha\beta\gamma} z_\alpha r_\beta
\; \bar U \gamma_\gamma \gamma_5 U ,
\label{bilinear_identity_1}
\\
\frac{1}{2} \, \bar U \left[ (rz) \gamma_\sigma - r_\sigma \hat{z} 
\right] U &=& 
i \epsilon_{\sigma\alpha\beta\gamma} z_\alpha p_\beta
\; \bar U \gamma_\gamma \gamma_5 U 
\label{bilinear_identity_2}
\ee
(similar identities hold with $\gamma$ replaced by $\gamma\gamma_5$ 
everywhere). Furthermore, to simplify the third term, we perform an
integration by parts in the piece with $\gamma_\gamma$ and make repeated
use of Eqs.(\ref{bilinear_identity_1}) and (\ref{bilinear_identity_2}):
\be
&& \frac{1}{2} \epsilon_{\sigma\alpha\beta\gamma} z_\alpha
r_\beta \; \bar U 
\left[ (pz) \gamma_\gamma - \hat{z} p_\gamma \right] 
\gamma_5 U 
\nonumber \\
&=& \frac{i}{4} 
\epsilon_{\sigma\alpha\beta\gamma} \epsilon_{\gamma\delta\epsilon\zeta} 
z_\alpha z_\delta r_\beta r_\epsilon \; \bar U \gamma_\zeta U 
\nonumber \\
&=& - \frac{i}{4} (rz) \, 
\bar U \left[ (rz) \gamma_\nu - r_\nu \hat{z} 
\right] U \; + \; \mbox{terms $\propto r^2$} 
\nonumber \\
&=& \frac{1}{2} (rz) \epsilon_{\sigma\alpha\beta\gamma} z_\alpha p_\beta 
\; \bar U \gamma_\gamma \gamma_5 U .
\label{bilinear_identity_3}
\ee
Using these identities one can easily convince oneself that the
three terms in the square bracket in Eq.(\ref{vector_quark}) add 
up to zero. The result for the vector string operator including 
symmetrized operators and total derivatives is thus, up to terms of 
order $r^2 = t$ which were dropped in Eq.(\ref{bilinear_identity_3}), 
completely given by:
\be
\langle p - r/2| \left[ \bar\psi (-z/2) \gamma_\sigma \psi (z/2) 
\right]^{\mbox{\scriptsize sym + total der.}}
| p + r/2 \rangle
&=& e^{-i (pz)} \; \bar U \gamma_\sigma U , 
\label{vector_quark_trivial}
\ee
which coincides with the matrix element of the vector string 
operator between free quark states. A similar result is found for the
axial vector string operator:
\be
\langle p - r/2| \left[ \bar\psi (-z/2) \gamma_\sigma \gamma_5 \psi (z/2) 
\right]^{\mbox{\scriptsize sym + total der.}} 
| p + r/2 \rangle
&=& e^{-i (pz)} \; \bar U \gamma_\sigma \gamma_5 U . 
\label{axial_quark_trivial}
\ee
\par
{\it Compton amplitude for the free quark.}
In particular, Eqs.(\ref{vector_quark_trivial}) and 
(\ref{axial_quark_trivial}) imply that the Compton amplitude for 
a free quark, calculated from the transverse generalization of the
leading term in the light--cone expansion,
Eq.(\ref{handbag_transverse}), is up to terms of order $t/q^2$
identical to the ``handbag'' Compton amplitude:
\beq
\int d^4 z \, e^{i (qz)} 
\langle p - r/2| \Pi_{\mu\nu}^{\mbox{\scriptsize transverse}} 
| p + r/2 \rangle
\;\; = \;\; \bar U \left[ 
\frac{\gamma_\mu (\hat p - \hat q) \gamma_\nu}
{(p - q)^2 + i0} \; + \; (q \rightarrow - q, \mu \leftrightarrow \nu) 
\right] U .
\eeq
This amplitude is known to be transverse, by virtue of 
Eqs.(\ref{Dirac_equation_quark}). To summarize, our calculation
shows that the operators dropped in Eq.(\ref{handbag_transverse})
indeed contribute to the free quark Compton amplitude only at 
power--suppressed level.
\par
It is instructive to take a look at the contributions of the individual
terms under the parameter integral in Eq.(\ref{vector_quark}) to the 
Compton amplitude. In the first and second term in the bracket, the 
sine and cosine functions of argument $\bar v (rz) / 2$ can be combined
with the overall exponential factor to give exponentials of the form
\beq
e^{-i [ v (pz) \mp \bar v (rz)/2 ]} .
\eeq
The contribution of these terms to the virtual Compton amplitude then
leads to integrals of the form
\beq
\int_0^1 dv \, v \, \int d^4 z \, e^{i (qz)} 
\left[ \frac{2 i z_\rho z_\alpha}{\pi^2 (z^2 - i0)^2} 
e^{-i [ v (pz) \mp \bar v (rz)/2 ]}  \right]
\;\; = \;\; \int_0^1 dv \, 
\frac{v}{[(q - v p \pm \bar v r/2)^2 + i0]^2}
\times \mbox{tensor} .
\eeq
In the generalized Bjorken limit these denominators become (up to terms of 
order $t/q^2$)
\beq
(q - v p \pm \bar v r/2)^2 
\;\; = 2 (p q) (-v \pm \bar v \eta - \xi ) ,
\eeq
where $\xi$ and $\eta$ are defined in Eq.(\ref{xi_eta_def}). In the DVCS 
limit $\eta = \xi$, and thus
\beq
(q - v p \pm \bar v r/2)^2 \;\; = \;\; 2 (p q) \left\{ 
\begin{array}{l} -v (1 + \xi ) \\[1.5ex] -2\xi - v (1 - \xi ) 
\end{array} 
\right.
\label{denominator_DVCS}
\eeq
In the case of the upper sign one is led to integrals of the form
\beq
\int_0^1 \frac{dv}{v + i0} ,
\eeq
which diverge logarithmically. This singularity is canceled
by a corresponding singularity produced 
by the third term in the bracket in Eq.(\ref{vector_quark}), which 
represents the contribution of the twist--2 operator with Dirac matrix
$\hat z \gamma_5$, as we have shown above, using the equations of motion 
for the quark spinors. As we shall see in Section~\ref{subsec_amplitude}, 
in the case of hadronic matrix elements, where the quarks are not on 
mass shell, this cancellation takes place only imperfectly, and a 
real divergence occurs in the Compton amplitude for transverse
photon polarization.
\section{DVCS amplitude including kinematical twist--3 terms}
\label{sec_amplitude}
\subsection{Spectral representation of twist--2 matrix elements}
\label{subsec_spectral}
We now use our result for the transverse generalization of the 
leading term in light--cone expansion of the product of 
electromagnetic currents to compute the DVCS amplitude for a hadronic
target. In this way we shall obtain an approximation for the DVCS
amplitude which is manifestly transverse up to terms of order 
$t/q^2$ or $m^2/q^2$.
\par
In order to compute the hadronic Compton amplitude we need to supply 
parametrizations of the hadronic matrix elements of the basic symmetrized
(twist--2) string operators in Eq.(\ref{handbag_transverse}). For simplicity
we restrict ourselves to the case of a pion target, which has spin zero
and simple charge conjugation properties. Specifically, we shall
consider the isoscalar component of the DVCS amplitude for the pion,
and suppress the overall factor resulting from the quark charges.
We shall neglect the pion mass and put $m = 0$.
The generalization of the expressions presented below to higher--spin targets
or to other isospin components of the amplitude is straightforward.
\par
{\it $z^2$--dependence. }
When computing the Compton amplitude, Eq.(\ref{Compton_def}), 
from the light--cone expansion in coordinate space, the integral
over the separation of the fields, $z$, runs over the whole 4--dimensional
space. Thus, in principle we need to provide a parametrization of 
the matrix elements valid everywhere in $z$, not only on 
the light cone. Consider the hadronic matrix element of the contracted
string operator, Eq.(\ref{string_sym}), for a massless spin--0 hadron. 
On general grounds it can be regarded as a 
function of the variables $pz, rz$ and $tz^2$:
\beq
\langle p - r/2 | \bar\psi (-z/2) \hat z \psi (z/2) | p + r/2 \rangle
\;\; \equiv \;\; \mbox{function}\, (pz, rz; t z^2) .
\eeq 
The dependence on $z^2$ is through the combination $t z^2$; the reason 
being that $t$ is the only dimensionful invariant in the problem.
Since we shall drop terms of order $t/q^2$ in the Compton amplitude,
it follows that we can effectively neglect the $z^2$--dependence
of the matrix element in the integral over $z$. In this sense
it is sufficient to provide a parametrization of the matrix elements 
of the operators only on the light cone, $z^2 = 0$. 
\par
One would have to take into account the $z^2$--dependence of the matrix 
elements if one wanted to compute kinematical corrections 
to the DVCS amplitude of order $t/q^2$ or $m^2/q^2$.
Such corrections are analogous to the well--known target mass
corrections to DIS \cite{Nachtmann:1973mr,Georgi:1976ve}. 
In this case one would have to
remember that the definition of the twist--2 part of the contracted operator
implies that it satisfies the harmonic condition 
[{\it cf.} Eq.(\ref{harmonic_O})]
\beq
\Box_z \left[ \bar\psi (-z/2) \hat z \psi (z/2) \right]^{\rm twist-2}
\;\; = \;\; 0 ,
\label{harmonic_me}
\eeq
which is equivalent to the condition that the local operators in the Taylor 
expansion in $z$ be traceless. Eq.(\ref{harmonic_me}) allows to reconstruct 
the $z^2$--dependence of the matrix element of the operator from ``initial
conditions'' given on the light--cone, $z^2 = 0$. In our calculation
here we shall explicitly put $t$ and $m^2$ to zero, so we do not
need to worry about trace subtraction. In this sense we do not
need to distinguish between the symmetrized string operator 
$\bar\psi (-z/2) \hat z \psi (z/2)$ and its twist--2 part.
\par
{\it Spectral representation for the matrix elements.}
The most convenient way to parametrize the hadronic matrix elements of the 
symmetrized string operators is by way of a spectral 
representation,
{\it i.e.}, a decomposition in plane waves. In the case of
the pion the matrix element of the pseudoscalar operator, 
$\bar\psi (-z/2) \hat z \gamma_5 \psi (z/2)$, is identically
zero. (The forward matrix element of this operator would be parametrized 
by the polarized parton density.) Thus we need only the 
parametrization of the matrix element of the contracted scalar 
twist-2 operator. For the isoscalar component of the matrix element
it is of the form
\be 
\lefteqn{
\langle p - r/2 | \bar\psi (-z/2) \hat z \psi (z/2) | p + r/2 \rangle }
&& \nonumber \\
&=& 2(pz) \, \int_{-1}^1 d\tilde x  
\int_{-1 + |\tilde x|}^{1-|\tilde x|} d \alpha  \; e^{-i (kz)} \,
f(\tilde x,\alpha)
\nonumber \\ 
&+& (rz) \, \int_{-1}^1 d \alpha \; e^{-i \alpha (rz)/2}\,
D(\alpha)
\nonumber \\[1ex]
&+&  \mbox{$z^2$--dependent terms} ,
\label{para1} 
\ee
where 
\beq
k \;\; = \;\; \tilde x p + \frac{\alpha}{2} r .
\eeq
In the partonic language, $k$ is the average of the outgoing, 
$\tilde xp + (1+\alpha) r/2$, and incoming, 
$\tilde  xP  - (1- \alpha) r/2$, parton momenta.
Here, $f(\tilde x,\alpha)$ is the so--called double 
distribution 
\cite{Radyushkin:1997ki,Radyushkin:1999es,Radyushkin:1999bz},
 while $D(\alpha)$ is the Polyakov-Weiss ({\rm PW}) 
distribution amplitude
absorbing the $(pz)$-independent terms of the matrix 
element \cite{Polyakov:1999gs}.
They obey the symmetry relations
\be
f(\tilde x , \alpha ) &=& -f(-\tilde x , \alpha) ,
\label{f_pion}
\\[1ex]
f(x , \alpha ) &=& \phantom{-} f(x , -\alpha ) , 
\label{f_Muenchen}
\\[1ex]
D(\alpha )     &=&  -D(-\alpha) .
\label{D_symm}
\ee
Eq.(\ref{f_Muenchen}) is the general symmetry of double distributions 
noted in Ref.\cite{MPW97}, while Eq.(\ref{f_pion}) is specific to the
their $C$-even components.
The $z^2$--dependent terms in Eq.(\ref{para1}) are in principle calculable 
in terms of $f(\tilde x,\alpha)$ and $D(\alpha )$ by solving the harmonic 
condition, Eq.(\ref{harmonic_me}), to all orders in $z^2$. However, since
each $z^2$ is accompanied by a factor of either $t$ or $m^2$,
we can neglect them, as explained above. 
\subsection{Matrix elements of vector--type string operators}
\label{subsec_vector}
From the parametrization of the matrix element of the contracted 
string operator, Eq.(\ref{para1}), we can obtain the matrix elements
of the the vector and axial vector string operators, 
Eq.(\ref{string_deconstructed_scalar}), including the kinematical
twist--3 contributions represented by the finite translations.
Let us consider first the part of the matrix elements coming from the double 
distribution term in Eq.(\ref{para1}); the contributions from the PW--term can
be included later. Substituting Eq.(\ref{para1})
into Eq.(\ref{string_deconstructed_scalar}), and using the fact that
under the matrix element the total derivative of string operators
turns into the momentum transfer,
\beq
\langle p - r/2 | i\frac{\partial}{\partial X_{\sigma}} 
\ldots | p + r/2 \rangle \;\; = \;\; 
r_{\sigma} \, \langle p - r/2 | \ldots | p + r/2 \rangle ,
\eeq
we obtain: 
\be 
\lefteqn{
\langle p - r/2 | 
\left[ \bar\psi (-z/2) \gamma_\sigma \psi (z/2) 
\right]^{\mbox{\scriptsize sym + total der.}} | p + r/2 \rangle }
&& \nonumber \\[1.5ex]
&=& \int_{-1}^1 d\tilde x  
\int_{-1 + |\tilde x|}^{1-|\tilde x|} d \alpha \, f(\tilde x,\alpha) 
\int_0^1 dv \, e^{-i v(kz)} 
\nonumber \\
&& \times \left\{  
2 \left[ p_{\sigma}  -ivk_{\sigma}(pz) \right] 
\cos [\bar v (rz)/2]
\; + \; v r_{\sigma} (pz) \sin  [\bar v (rz)/2]
\right\} 
\nonumber 
\\[1.5ex]
&=& \int_{-1}^1 d\tilde x  
\int_{-1 + |\tilde x|}^{1-|\tilde x|} d\alpha \,f(\tilde x,\alpha) 
\, \left( 2 p_{\sigma} e^{-i (kz)} \phantom{\int}
\right.
\nonumber \\
&-& \left. \int_0^1 dv \, v \, e^{-i v(kz)} 
\left\{ \sin  [\bar v (rz)/2]
\left[ p_{\sigma} (rz) - r_{\sigma} (pz) \right] 
- 2 i \cos [\bar v (rz)/2]
\left[ p_{\sigma}(kz) - k_{\sigma} (pz) \right]  
\right\} \right) .
\label{para3} 
\ee
In the last step we have performed an integration by parts in the integral
over the parameter $v$ in the term proportional to $p_{\sigma}$.
As described in Section~\ref{sec_virtual}, in DVCS kinematics we
can express the momentum transfer in the form 
[{\it cf.}\ Eq.(\ref{r_thru_Delta})]
\beq
r \;\; = \;\; 2 \xi p + \Delta ,
\eeq
where $\Delta$ is of order $|t|^{1/2}$.
In the partonic language the term proportional to $p$ would be the
longitudinal component, and $\Delta$ the transverse component. 
Similarly, we write the ``active'' quark momentum in the form
$k=(\tilde x+\xi \alpha)p + \alpha \Delta / 2$. 
Noting that in the square brackets in Eq.(\ref{para3})  
only the $\Delta$--parts of the momenta $r$ and $k$ contribute,
\be 
p_{\sigma} (rz) - r_{\sigma} (pz) &=& 
\;\;\;\; p_{\sigma} (\Delta z) - \Delta_{\sigma} (pz) ,
\nonumber \\
p_{\sigma}(kz) - k_{\sigma} (pz) &=& 
\frac{\alpha}{2}
\left[ p_{\sigma}(\Delta z) - \Delta_{\sigma} (pz) \right]
\ee
we obtain for the matrix element
\be 
\lefteqn{
\langle p - r/2\, | \left[ \bar\psi (-z/2) \gamma_\sigma \psi (z/2) 
\right]^{\mbox{\scriptsize sym + total der.}} 
| \, p +r/2  \rangle } && \nonumber 
\\[1.5ex]
&=&  \int_{-1}^1 d \tilde x \int_{-1 + |\tilde x|}^{1-|\tilde x|} 
d \alpha \, f(\tilde x,\alpha)  \, \left(
2 p_{\sigma}
e^{-i (\tilde x+ \xi \alpha )(pz) -i\alpha (\Delta z)/2 }  
\phantom{\int} \right.
\nonumber \\ 
&+& \left. \left[ \Delta_{\sigma} (pz) - p_{\sigma}(\Delta z) \right]
\int_0^1 dv \, v\, 
e^{-i v (\tilde x + \xi \alpha )(pz)-iv\alpha (\Delta z)/2 } 
\, \left\{  \sin  [\bar v (rz)/2] 
- i \alpha  \cos [\bar v   (rz)/2]  \right\} 
\right) . 
\label{para4} 
\ee
\par
\begin{figure}[t]
\begin{center}
\setlength{\epsfxsize}{7.5cm}
\setlength{\epsfysize}{7.5cm}
\mbox{\epsffile{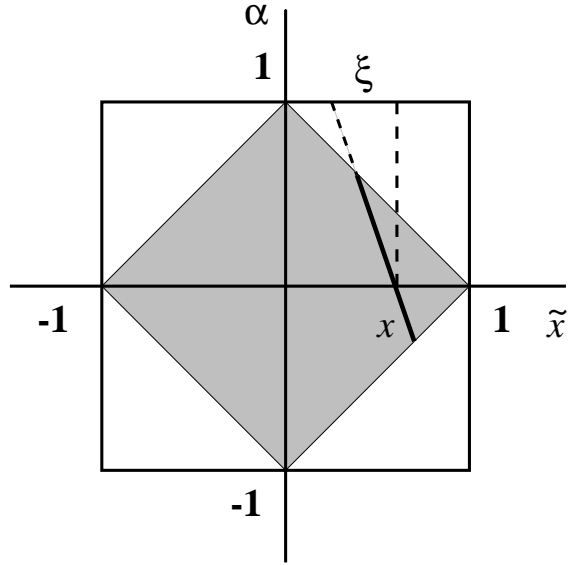}}
\end{center}
\caption[]{Graphical representation of the reduction formulas, 
Eq.(\ref{reduction}), defining the SPD's, $H(x, \xi)$ and $A(x, \xi)$, 
in terms of the double distribution, $f(\tilde x, \alpha )$.
The range of the arguments $\tilde x$ and $\alpha$ 
of the double distribution is the ``rhombus''
defined by $|\tilde x| + |\alpha | \le 1$.
The integration is over the line defined by the
condition $\tilde x  +  \xi \alpha = x$ and lying inside the 
range of the variables $\tilde x, \alpha$; it is indicated by the
fat line. The variable $x$ defines the intersection of this line with
the $\tilde x$--axis, the variable $\xi$ its deviation from the
vertical. For $x = \xi$ the line runs through the upper corner
of the rhombus.}
\label{fig_reduc}
\end{figure}

{\it Introducing skewed distributions.}
Since we aim to compute the DVCS amplitude up to terms of order $t/q^2$, we
can expand the matrix elements of the vector string operator in 
$\Delta \sim |t|^{1/2}$. This will allow us to express the matrix elements 
in terms of ``skewed'' distributions, {\it i.e.}, one--variable spectral functions
depending on the kinematical variable $\xi$ as an external
parameter. Expanding the exponential factors and the sine and cosine
functions in Eq.(\ref{para4}) in $(\Delta z)$ we obtain, to first order in
$\Delta$:\footnote{Since we plan to drop terms of order $t/q^2$ in the
Compton amplitude it seems sufficient to expand the matrix elements to
first order in $\Delta$, since $\Delta^2$--terms would already be
proportional to $t$.  There is, however, a subtle point concerning
contributions to the Compton amplitude of tensor structure 
$\Delta_\mu \Delta_\nu$, which one needs to include in order to convert an
approximately gauge--invariant structure in the tensor amplitude into an
exactly gauge--invariant one. Although the $\Delta_\mu \Delta_\nu$ term is
formally of order $O(\Delta^2 )$, it can be obtained only from the
contribution corresponding to the leading light--cone singularity, 
$\hat z /(z^2)^2$, {\it cf.}\ the discussion at the end of
Section~\ref{subsec_amplitude}.}
\be 
\lefteqn{
\langle p - r/2 | \left[
\bar\psi (-z/2) \gamma_\sigma \psi (z/2) 
\right]^{\mbox{\scriptsize sym + total der.}} 
| p + r/2 \rangle }
&& \nonumber \\
&=& \int_{-1}^1 d \tilde x \int_{-1 + |\tilde x|}^{1-|\tilde x|} 
d \alpha \, f(\tilde x,\alpha)  \, \left(
2 p_{\sigma} \left[ 1 - \frac{i\alpha}{2} (\Delta z) \right] 
e^{-i (\tilde x+ \xi \alpha )(pz) } \right.
\nonumber \\  
&+& \left. \left[ \Delta_{\sigma} (pz) - p_{\sigma}(\Delta z) \right]
\int_0^1 dv \, v\, e^{-i v (\tilde x + \xi \alpha )(pz)} 
\, \left\{  \sin [\bar v \xi(pz)]
- i  \alpha  \cos [\bar v  \xi (pz)]  \right\} \right) .
\label{para5} 
\ee
Now the spectral parameters, $\tilde x$ and $\alpha$, appear in 
the exponential factors only in the combination 
\[
x \;\; \equiv \;\; \tilde x + \xi \alpha ,
\]
so the information entering in the matrix element is effectively
contained in certain one--dimensional ``reductions'' of the double 
distribution, $f(\tilde x, \alpha)$. We introduce two skewed parton 
distributions:
\beq
\left.
\begin{array}{r}
H(x, \xi) \\[1.5ex]
A(x, \xi)
\end{array}
\right\}
\;\; \equiv \;\; \int_{-1}^1 d\tilde x  
\int_{-1 + |\tilde x|}^{1-|\tilde x|}  d \alpha   \, 
\delta (x  - \tilde x  -  \xi \alpha ) \, f(\tilde x, \alpha) 
\; \left\{
\begin{array}{r}
1 \\[1.5ex]
\alpha
\end{array}
\right. .
\label{reduction}
\eeq
This is visualized in Fig.~\ref{fig_reduc}.
%
%
As a consequence of the symmetries of the double distribution,
Eqs.(\ref{f_pion}) and (\ref{f_Muenchen}), the functions $H$ and $A$ 
satisfy the symmetry relations
\beq
\begin{array}{rcrrcr}
H(x , \xi) &=& -H(-x , \xi) , & \hspace{5em} H(x , \xi) &=&  H(x , -\xi) , 
\\[2ex]
A(x , \xi) &=&  A(-x , \xi) , & A(x , \xi) &=& -A(x , -\xi) .
\end{array}
\label{H_D_symm}
\eeq
Furthermore, because of the antisymmetry of the combination 
$ \alpha f (\tilde x, \alpha)$  
with respect both to
$x$ and $\alpha$ we have
\beq 
\int_{0}^1 dx \, A (x, \xi) \;\; = \;\; 0 .
\eeq
Note that the integral over $x$ is taken  
over the $0 \leq x \leq 1$ interval.    
Hence, the distribution $A(x, \xi)$ 
cannot be a  positive-definite function  
on $0 \leq x \leq 1$.
\par
Combining in Eq.(\ref{para5}) similar terms we obtain the 
following representation of the matrix element in terms 
of the skewed parton distributions: 
\be 
\lefteqn{
\langle p - r/2 | \left[ 
\bar\psi (-z/2) \gamma_\sigma \psi (z/2) 
\right]^{\mbox{\scriptsize sym + total der.}} 
| p + r/2 \rangle }
&& \nonumber \\
&=& \int_{-1}^1 dx \left( 2 p_{\sigma} 
 e^{-i x(pz) } \left[ H(x, \xi) - \frac{i (\Delta z)}{2} 
A(x, \xi) \right] \phantom{\int} \right. 
\nonumber \\ 
&+& \left.
\left[ \Delta_{\sigma} (pz)  - p_{\sigma}(\Delta z) \right]
\int_0^1 dv \, v \, e^{-i v x (pz)} 
\left\{  H(x, \xi)\, \sin [\bar v \xi(pz)]
- i A(x, \xi) \, \cos [\bar v \xi (pz)]
\right\} \right) .
\label{para6} 
\ee
The cosine and sine functions can be represented as 
the sum or difference of exponentials. Combining them 
with the overall exponential factor, $e^{-i vx (pz)}$,
one gets $vx \pm \bar v \xi$ combinations.
However, using the symmetries in $x$ of the SPD's, 
Eq.(\ref{H_D_symm}), one can  easily arrange that only $vx + \bar v \xi$
appears. This finally gives 
\be 
\lefteqn{
\langle p - r/2 | \left[
\bar\psi (-z/2) \gamma_\sigma \psi (z/2) 
\right]^{\mbox{\scriptsize sym + total der.}} 
| p + r/2 \rangle }
&& \nonumber \\
&=&  
\int_{-1}^1 d x  \left( 2 p_{\sigma} 
e^{-i x(pz) } \left[ H(x, \xi) 
-  \frac{i (\Delta z)}{2} A(x, \xi) \right] \phantom{\int} \right. 
\nonumber \\ 
&+& \left.
\frac{i}{2} \, \left[ \Delta_{\sigma} (pz)  - p_{\sigma}(\Delta z) \right]
\int_0^1 dv \, v \, 
 \left[ e^{-i (v x +\bar v \xi) (pz)}
+ e^{i (v x +\bar v \xi) (pz)} \right] 
\left[ H(x, \xi) - A(x, \xi) \right] \right) .
\label{para7} 
\ee
\par
{\it The axial vector string operator.}
In a similar fashion we compute the matrix element of the axial vector
string operator, Eq.(\ref{string_deconstructed_scalar}).
Substituting the double distribution part of the spectral representation
of the contracted operator, Eq.(\ref{para1}), into 
Eq.(\ref{string_deconstructed_scalar}), we obtain
\be 
\lefteqn{
\langle p - r/2 | \left[
\bar\psi (-z/2) \gamma_\sigma \gamma_5 
\psi (z/2) \right]^{\mbox{\scriptsize sym + total der.}} 
| p + r/2 \rangle }
&& \nonumber \\
& = & 2 \epsilon_{\sigma\alpha\beta\gamma} z_\alpha 
r_{\beta}  
\int_{-1}^1 d\tilde x  
\int_{-1 + |\tilde x|}^{1-|\tilde x|} d \alpha \, f(\tilde x,\alpha) \,
\int_0^1 dv  
\left[ p_{\gamma} - ivk_{\gamma} (pz) \right] \, e^{-i v(kz)}\,
\frac{\sin [\bar v (rz)/2]}{(rz)} .
\label{parax1} 
\ee
Again integrating by parts in the $p_{\gamma}$ term, and keeping only 
terms linear in the transverse component of the momentum transfer, 
$\Delta$, this turns into
\be 
\lefteqn{
\langle p - r/2 | \left[ \bar\psi (-z/2) \gamma_\sigma \gamma_5 
\psi (z/2) \right]^{\mbox{\scriptsize sym + total der.}} 
| p + r/2 \rangle }
&& \nonumber \\
& = & \epsilon_{\sigma\alpha\beta\gamma} \, z_\alpha 
 \Delta_{\beta}    p_{\gamma}
\int_{-1}^1 d\tilde x  
\int_{-1 + |\tilde x|}^{1-|\tilde x|} d \alpha \,f(\tilde x,\alpha)
    \int_0^1 dv \, v \, e^{-i v(\tilde x+\xi \alpha)(pz) } 
\left\{    \cos [\bar v \xi (pz)]
+ {i\alpha} \, \sin [\bar v \xi (pz)] \right\} .
\label{parax3} 
\ee
One sees that this matrix element can be expressed in terms of the
same skewed distributions, $H$ and $A$, as the one of the vector 
operator:
\be 
\lefteqn{
\langle p - r/2 | \left[ \bar\psi (-z/2) \gamma_\sigma \gamma_5 
\psi (z/2) \right]^{\mbox{\scriptsize sym + total der.}} 
| p + r/2 \rangle }
&& \nonumber \\
&=& 
\frac{1}{2} \epsilon_{\sigma\alpha\beta\gamma} \, z_\alpha 
\Delta_{\beta} p_{\gamma}
\int_{-1}^1 dx 
\int_0^1 dv \, v \, 
\left[ e^{-i (v x +\bar v \xi) (pz)}
- e^{i (v x +\bar v \xi) (pz)} \right] 
\left[ H(x, \xi) - A(x, \xi) \right] .  
\label{parax5} 
\ee
\par
Hence, the matrix elements of both the vector and axial vector string 
operators, including the twist--2 and kinematical 
twist--3 contributions, are completely described by the two skewed 
distributions, $H$ and $A$, which, in turn, are determined by the
original double distribution, Eq.(\ref{para1}), through the
reduction formulas, Eq.(\ref{reduction}).
\par
{\it Contribution from the PW--term.} In order to obtain the full matrix 
elements of the vector and axial vector operator we must add to 
Eqs.(\ref{para7}) and (\ref{parax5}) the parts coming from the PW--term 
in the spectral representation of the matrix
element of the contracted operator, Eq.(\ref{para1}). Substituting
this term in Eq.(\ref{string_deconstructed_scalar}), and going through
the same steps as described above for the double--distribution part,
one easily sees that the contribution of the PW term to the
matrix element of the vector operator is of the form
\be 
\langle p - r/2 | \left[
\bar\psi (-z/2) \gamma_\sigma \psi (z/2) 
\right]^{\mbox{\scriptsize sym + total der.}} 
| p + r/2 \rangle_{\mbox{\scriptsize PW--Term}}
\;\; = \;\;
r_{\sigma}  \int_{-1}^1 d \alpha  \, e^{-i \alpha (rz)/2} \, 
D(\alpha) .
\label{para_PW} 
\ee
Note that the {\rm PW}-term has a simple structure, corresponding to a 
parton picture in which the partons carry fractions $(1 \pm \alpha)/2$ 
of the momentum transfer, $r$. Since only one momentum, $r$, is 
involved, this term can contribute only to the totally symmetric part
of the vector string operator, and thus ``decouples'' in the 
reduction relations, Eqs.(\ref{rest_5_total_derivative}) 
and (\ref{rest_5_total_derivative}). For the same reason the
{\rm PW} term also does not contribute to the matrix element of 
the axial vector string operator: The derivatives with respect to 
both coordinates, $X$ and $z$, give rise to the momentum transfer, 
$r$, whence the contraction with the $\epsilon$--tensor gives zero.
\subsection{DVCS amplitude for pion target}
\label{subsec_amplitude}
Having at hands the parametrizations of the matrix elements of the 
vector and axial vector string operators 
we can now compute the DVCS amplitude for a pion target.
Substituting the parametrizations, Eqs.(\ref{para7}) and (\ref{parax5}),
into Eq.(\ref{handbag_transverse}),
and performing the Fourier integral over the distance, $z$, we
obtain the hadronic Compton amplitude in the form
\be
T_{\mu\nu} &=& \int_0^1 dx \int d^4 z \; e^{i (q z)} 
\frac{2 i z_\rho}{\pi^2 (z^2 - i0)^2}
\nonumber \\
&\times& \left\{ 2 s_{\mu\rho\nu\sigma} p_\sigma 
\; e^{-i x(pz) } \; \left[ H(x, \xi) 
-  \frac{i (\Delta z)}{2} A(x, \xi) \right]
\right.
\nonumber \\ 
&& + \; \frac{i}{2} \, s_{\mu\rho\nu\sigma} 
\left[ \Delta_{\sigma} (pz)  - p_{\sigma}(\Delta z) \right]
\nonumber \\
&& \;\;\;\;\; \times \int_0^1 dv \, v \, 
 \left[ e^{-i (v x +\bar v \xi) (pz)}
+ e^{i (v x +\bar v \xi) (pz)} \right]
\left[ H (x, \xi) - A(x, \xi) \right]
\nonumber \\
&& - \; \frac{i}{2}\,  \epsilon_{\mu\rho\nu\sigma}
\epsilon_{\sigma\alpha\beta\gamma} \, z_\alpha 
\Delta_{\beta} p_{\gamma}
\nonumber \\
&& \left. \;\;\;\;\; \times 
\int_0^1 dv \, v \, 
 \left[ e^{-i (v x +\bar v \xi) (pz)}
- e^{i (v x +\bar v \xi) (pz)} \right]
\left[ H (x, \xi) - A(x, \xi) \right]
\right\} 
\nonumber
\\
&+& (q \rightarrow -q, \mu \leftrightarrow \nu ) .
\ee
Using the well--known identity for the contraction of two 
$\epsilon$--tensors one sees that the second and third term 
here can be combined to give two terms involving only 
exponential factors with positive and negative sign, respectively,
which give identical contributions.
The integral over $z$ is readily computed,
using the basic formulas:
\be
\int d^4 z \,  e^{i(lz)} \frac{z_\rho}{2 \pi^2 (z^2 - i0)^2}
&=& \frac{l_\rho}{l^2 + i0} ,
\\
\int d^4 z \, e^{i(lz)} \frac{i z_\rho z_\alpha}{2 \pi^2 (z^2 - i0)^2}
&=& \frac{g_{\rho\alpha} l^2 - 2 l_\rho l_\alpha}{(l^2 + i0 )^2} .
\ee
Here $l$ is to be replaced by the sum of all momenta appearing in the 
exponential factors, and $l^2$ in the denominators can be simplified
using the special relations between invariants in DVCS kinematics,
{\it cf.}\ Section~\ref{sec_virtual}. Making use of the symmetry properties 
of the pion SPD's, $H$ and $A$, Eq.(\ref{H_D_symm}), the result can be 
assembled in the form 
\be
T_{\mu\nu} &=& \frac{1}{2(pq)} \left\{ \;
2 \left[ p_\mu q_\nu + q_\mu p_\nu - g_{\mu\nu} (pq) + 2 \xi p_\mu p_\nu
\right] \int_{-1}^1 dx \frac{H(x, \xi)}{x - \xi + i0} 
\right.
\nonumber
\\
&& \phantom{\frac{1}{2(pq)}} 
+ \; \left[ \Delta_\mu p_\nu + \Delta_\nu p_\mu \right]
\int_{-1}^1 dx \frac{A(x, \xi)}{x - \xi + i0} 
\nonumber
\\
&& \phantom{\frac{1}{(pq)}} 
+ \; \int_{-1}^1 dx [H(x, \xi) - A(x, \xi)]
\nonumber
\\
&& \phantom{\frac{1}{2(pq)}} \;\;\;\;\; \left. \times \int_0^1 dv \, v \, 
\left[
\frac{q_\mu - (v x + \bar v \xi ) p_\mu }
{(\xi + v x + \bar v \xi - i0 )^2}  \Delta_\nu
\; + \; \Delta_\mu \frac{q_\nu + (v x + \bar v \xi ) p_\nu}
{(-\xi + v x + \bar v \xi - i0)^2}
\right] \right\} .
\label{Compton_bare}
\ee
We can bring this result into a form which is manifestly
transverse with respect to both the incoming and outgoing photon
momenta. By simple algebraic rearrangement, and making use of the 
integral identity
\be
\int_0^1 dv \, v \, \frac{- v x - \bar v \xi}
{(\xi + v x + \bar v \xi - i0)^2}
+ \frac{1}{x + \xi - i0} 
&=& \int_0^1 dv \, v \, \frac{3 \xi}
{(\xi + v x + \bar v \xi - i0)^2} ,
\label{id_Delta_mu}
\ee
which holds if $x \ne \pm \xi$, we first write Eq.(\ref{Compton_bare})
as a sum of three terms which are individually 
transverse \cite{Anikin:2000em}. In addition, we then perform an integration 
by parts in $x$ in order to remove the quadratic denominators. Here we use
that
\beq
H(x, \xi) - A(x, \xi) \;\; = \;\; 0 \;\;\;\;\; \mbox{for} \;\;\;\;\;
x \;\; = \;\; \pm 1 ,
\eeq
which naturally follows from the reduction formulas, Eq.(\ref{reduction}).
In this way we finally obtain
\be
T_{\mu\nu} &=& \frac{1}{2(pq)} \left\{ \; 
2 \left[ p_\mu q_\nu + q_\mu p_\nu - g_{\mu\nu} (pq) + 2 \xi p_\mu p_\nu
-  p_\mu \frac{\Delta_\nu}{2}  + \frac{\Delta_\mu}{2} p_\nu   
\right] \int_{-1}^1 dx \frac{H(x, \xi)}{x - \xi + i0} 
\right.
\nonumber
\\[2ex]
&& \phantom{\frac{1}{2(pq)}} + \; (q_\mu + 3 \xi p_\mu) \Delta_\nu 
\int_{-1}^1 dx \int_0^1 dv \, \frac{R(x, \xi)}
{\xi + v x + \bar v \xi - i0}
\nonumber
\\[2ex]
&& \phantom{\frac{1}{2(pq)}} \left. + \; \Delta_\mu (q_\nu + \xi p_\nu) 
\int_{-1}^1 dx \int_0^1 dv \,  
\frac{R(x, \xi)}{-\xi + v x + \bar v \xi + i0} \;
\right\} ,
\label{Compton_Anikin}
\ee
where $R(x, \xi )$ is a short--hand notation for the following combination 
of derivatives of the SPD's:
\beq
R (x, \xi) \;\; \equiv \;\; \frac{\partial}{\partial x} H(x,\xi) 
- \frac{\partial}{\partial x} A(x,\xi) .
\label{R_def}
\eeq
All three terms here are individually transverse up to terms of order
$t$. The decomposition of the Compton amplitude here is identical to the
one introduced in Ref.\cite{Anikin:2000em}.  The first term is the twist--2
part with the tensor structure corrected as suggested by Guichon and
Vanderhaeghen \cite{Guichon}.  The second and third term represent twist--3
contributions.  Note that the second term contributes only to the helicity
amplitude for a longitudinally polarized initial (virtual) photon, the
third term for a transversely polarized one.\footnote{In general the virtual
Compton amplitude for a spin--0 target has five independent structures. The
fact that in the approximation considered here (twist--2 plus kinematical
twist--3, neglecting $t/q^2$ and $m^2/q^2$ contributions) only three of
these structures are non-zero is analogous to the absence of the
longitudinal structure function in DIS at twist--2 level due to the
Callan--Gross relation.}
\par
One observes that the second SPD, $A(x, \xi )$, enters in
Eq.(\ref{Compton_Anikin}) only through its first derivative, {\it cf.}\
Eq.(\ref{R_def}). From the reduction formulas, Eq.(\ref{reduction}), one
sees that the $x$--derivative of the function $A$ is related to the
$\xi$--derivative of $H$ by
\beq
\frac{\partial}{\partial x} A(x, \xi ) \;\; = \;\;
-\frac{\partial}{\partial \xi} H(x, \xi) ,
\eeq
so that instead of Eq.(\ref{R_def}) one could equivalently write
\beq
R (x, \xi) \;\; = \;\; \left( \frac{\partial}{\partial x}
+ \frac{\partial}{\partial \xi} \right) H(x,\xi) .
\label{R_as_xi_derivative}
\eeq
One could thus express the amplitude, Eq.(\ref{Compton_Anikin}), completely
in terms of the first SPD, $H$, and its $\xi$--derivative. In this way one 
would obtain the form of the Compton amplitude given in 
Ref.\cite{Kivel:2000rb}.
\par
{\it Inspecting the singular integrals.} We now inspect the singular
integrals which appear in the the expression for the Compton amplitude,
Eq.(\ref{Compton_Anikin}). The integrand in the first term has only a
simple pole at $x = \xi$; this integral is of the same type as that
obtained keeping only the twist--2 part of the matrix elements. The
integral exists if the SPD, $H$, is continuous at $x = \xi$, which is the 
case for SPD's derived from DD's which are less singular than $1/\tilde x^2$ 
for $\tilde x \rightarrow 0$ and otherwise 
continuous \cite{Musatov2000}.\footnote{Continuous SPD's were
also obtained in model calculations of the SPD's at a low scale in the
instanton vacuum \cite{Petrov:1998kf}.} The second and third term in
Eq.(\ref{Compton_Anikin}) involve additional parameter integrals, so their
convergence needs to be investigated carefully. In the second term, which
contributes to the helicity amplitude for longitudinal polarization, we can
perform the parameter integral over $v$, excluding the value $x = \xi$ for
the moment:
\beq
\int_0^1  
\frac{dv}{\xi + v x + \bar v \xi - i0}
\;\; = \;\;
\frac{1}{x - \xi } \ln \left( \frac{x + \xi - i0}{2 \xi - i0} \right) .
\eeq
This function has a logarithmic singularity at $x = - \xi$, while
it is regular at $x = \xi$. The integral over $x$ in Eq.(\ref{Compton_Anikin}) 
exists if $R(x, \xi)$ is bounded at $x = -\xi$, which again is the case
in the DD--based models described in  Ref.\cite{Musatov2000}.  
We thus obtain a finite answer also for the second term
contributing to longitudinal photon polarization.
\par
Let us now turn to the third term of Eq.(\ref{Compton_Anikin}). Here
we are dealing with an integral of the type (remember that 
$\bar v = 1 - v$)
\beq
\int_0^1 \frac{dv}{-\xi + v x + \bar v \xi + i0} 
\;\; = \;\;
\int_0^1  \frac{dv}{v (x - \xi) + i0} .
\label{divergent_integral}
\eeq
This integral exhibits a logarithmic divergence at the lower limit.
One may  hope to  get a finite result only if the integral 
\beq
I(\xi) \;\; \equiv \;\; \int_{-1}^1 dx
  \frac{R(x,\xi)}{x - \xi + i0} 
\label{Ixi} 
\eeq
vanished. However, one can easily convince oneself that
this quantity is generally non-zero. In fact, representing
$R(x, \xi )$ as in Eq.(\ref{R_as_xi_derivative}), and performing
an integration by parts in the term with the $x$--derivative,
one finds that
\beq
I(\xi) \;\; = \;\; \frac{d}{d\xi} 
\int_{-1}^1 dx \frac{H(x, \xi)}{x - \xi + i0} ,
\label{xider}  
\eeq
which is the $\xi$--derivative of the twist--2  
contribution to the DVCS amplitude. Since the latter generally has a 
nontrivial $\xi$--dependence, determined by the shape of the SPDs,
we conclude that the parameter integral Eq.(\ref{divergent_integral})
really is divergent. Thus, in our approximation the twist--3 part of the 
tensor amplitude $T_{\mu\nu}$ corresponding to transverse polarization of 
the initial photon contains a divergence. A similar observation has 
recently been made by Kivel {\it et al.}\ \cite{Kivel:2000rb}. 
\par
{\it Absence of divergences in the physical DVCS amplitude.}
One may wonder if the divergence occurring in the tensor amplitude for
transverse polarization of the initial photon affects the physical
DVCS amplitude. Fortunately, this is not so. One can easily see that
the tensor structure of the third term of Eq.(\ref{Compton_Anikin}) 
involving the divergent integral,
\beq
\Delta_\mu (q_\nu +\xi p_\nu) ,
\eeq
is just a truncated (to first order in $\Delta$) version of the exactly 
gauge invariant combination,
\beq
\Delta_\mu {q_2}_{\nu} 
\;\; = \;\;
\Delta_\mu (q_\nu + r_\nu / 2)
\;\; = \;\;
\Delta_\mu (q_\nu + \xi p_\nu + \Delta_\nu / 2) ,
\eeq
which has zero projection on the polarization vector of
the final real photon, $\epsilon_{2\nu}$: $(\epsilon_2 q_2) = 0$. 
In fact, the structure $\Delta_\mu {q_2}_{\nu}$ is what one would
obtain if one calculated the DVCS amplitude directly from the original 
``unexpanded'' form of the DD parametrization of the matrix element,
Eq.(\ref{para4}). It appears from the 
term with the exponential factor of the argument 
\beq
-i [v(\tilde x
+\xi \alpha)(pz)+ \bar v \xi (pz) + (v \alpha +\bar v) (\Delta z)/2] ,
\eeq
which is obtained by combining the sine/cosine functions and the
exponential factor in the second term of Eq.(\ref{para4}). In the
Compton amplitude, this term gives rise to a contribution in which the
argument of the quark propagator is 
\beq
q+v(\tilde x +\xi \alpha)p +
\bar v \xi p + (v \alpha + \bar v)\Delta/2 .  
\eeq
Since $(\Delta q), (\Delta p$) and $\Delta^2 $ are negligible, the 
denominator factors in Eq.(\ref{Compton_Anikin}) remain the same
as above. In the numerators, however, writing
$(v \alpha + \bar v)\Delta/2$ as $[1-(1-\alpha)v]\Delta/2$, 
we observe that the $\Delta_{\mu} (q_\nu+\xi
p_\nu)$ term of the above result converts into 
$\Delta_{\mu} (q_\nu+\xi p_\nu
+\Delta_\nu/2 ) = \Delta_\mu {q_2}_{\nu}$, plus a contribution
of the form $v \Delta_{\mu} \Delta_{\nu}$. The latter corresponds to a 
``new'' SPD defined by a reduction formula of the type of Eq.(\ref{reduction})
but with integrand $(1-\alpha)^2 f(\tilde x, \alpha)$. Due to the extra 
factor $v$, the $v$--integral for the latter contribution is finite. 
Thus, inclusion of the $\Delta_\mu \Delta_\nu$ term in the matrix elements
and the Compton amplitude relegates the divergence in the tensor amplitude
to the (exactly) unphysical part which does not contribute to the photon 
helicity amplitudes.
\par
It may appear that retaining the $\Delta_\mu \Delta_\nu$ term in the 
Compton amplitude would exceed the accuracy of our approximation, since
this term is formally of order $t$, and thus of the same order as
corrections from $z^2$--terms in the matrix elements which we did not 
take into account. However, the $\Delta_\mu \Delta_\nu$ term is distinct
in the sense that it can {\it only} arise in the way described above,
{\it i.e.}, from keeping finite $\Delta$ in the exponents of the 
matrix elements, while $z^2$--terms in the matrix elements can give
rise only to corrections of order $t/q^2$ to {\it already existing} tensor 
structures. In this sense the inclusion of $z^2$--terms would not
spoil the above argument proving that the physical Compton amplitude
is divergence--free up to terms of order $t/q^2$.
\par
To summarize, for the physical DVCS amplitude ({\it i.e.}, the photon
helicity amplitudes) we find no evidence for factorization breaking in 
the kinematical twist--3 contributions, both in the $\sqrt{t/q^2}$ 
and $t/{q^2}$ terms. It is quite possible that factorization breaks down 
at the $t/{q^2}$ level, but in order to establish this one would need 
to analyze the contributions from $z^2$--suppressed terms in the
matrix element ({\it i.e.}, twist--4 contributions).
\par
{\it Contribution from the PW--term.} The complete result for the
Compton amplitude includes also the contribution from the PW
term. Since this part of the matrix element of the contracted operator
``decouples'' and does not produce kinematical twist--3 terms,
its contribution to the Compton amplitude should be transverse by 
itself. Indeed, a straightforward calculation, 
making use of the symmetry Eq.(\ref{D_symm}), gives 
\beq
T_{\mu\nu}|_{\mbox{\scriptsize PW--Term}} \;\; = \;\; 
\frac{1}{(rq)}
\left[ r_{\mu} q_{\nu} +  q_{\mu} r_{\nu}  -  g_{\mu \nu} (rq)
+ r_{\mu} r_{\nu} \right] \; \int_{-1}^1 d \alpha \;
\frac{D(\alpha)}{1-\alpha} ,
\label{PW} 
\eeq
which evidently satisfies 
\beq
q_{\mu} T_{\mu \nu}|_{\mbox{\scriptsize PW--Term}} \;\; = \;\; 0, 
\hspace{5em}
r_{\mu} T_{\mu \nu}|_{\mbox{\scriptsize PW--Term}} \;\; = \;\; O(t/q^2 ).
\eeq
Hence this term can be regarded as a separate contribution.
Alternatively, one may include the contribution from the PW--term
into the basic SPD $H(x,\xi)$ and all SPD's derived from it.
Specifically, for $\xi >0$, the PW contribution 
to $H(x,\xi)$ is $D (x/\xi) \, \theta (|x| \leq \xi)$ 
\cite{Polyakov:1999gs}; 
it contributes   $(\xi-x)D'(x/\xi)\, \theta (|x| \leq \xi)/\xi^2$
[where $D'(\alpha) \equiv (d/d\alpha) D(\alpha)] $ to 
$R(x,\xi)$. 
Inserting these functions into Eqs. (\ref{Compton_Anikin}) and 
(\ref{comfin}) one rederives Eq. (\ref{PW}). 
One can also observe that the PW term makes zero contribution
to $I(\xi)$, Eq. (\ref{Ixi}).  
\subsection{Wandzura--Wilczek--type representation of twist--3 SPD's}
\label{subsec_WW}
We can express the results for the matrix elements and the Compton
amplitude derived in Sections~\ref{subsec_vector} and
\ref{subsec_amplitude} in an alternative form, introducing new skewed
distributions as Wandzura--Wilczek--type transformations of the
original SPD's, $H$ and $A$. This allows us to demonstrate that our
approach is equivalent to the Wandzura--Wilczek approximation proposed in
Ref.\cite{Belitsky:2000vx}. In this language, the singularity for transverse
photon polarization described in Section~\ref{subsec_amplitude} appears due
to discontinuities of the new SPD's at $x = \xi$.
\par
Within our approach the step leading to a Wandzura--Wilczek--type 
formulation is to introduce in the parameter integrals over $v$
the combination
\beq
v x + \bar v \xi
\eeq
as a new variable. With $R(x, \xi)$ given by Eq.(\ref{R_def}), we 
define a ``transformed'' function by way of 
\be 
R_W (x, \xi) &\equiv& 
\int_{-1}^1 dy \, R (y, \xi) 
\int_0^1 dv \, \delta ( yv + \bar v \xi -  x)
\nonumber \\
&=& \theta ( x \geq \xi) \int_{x}^1 dy \; 
\frac{R(y,\xi)}{y-\xi}  \; - \; \theta ( x \leq \xi) \int_{-1}^x 
dy \; \frac{R(y,\xi)}{y-\xi} .
\label{rw}
\ee
where $x \ne \xi$. Note that the function
\beq
R(x, \xi) \;\; = \;\; \frac{\partial}{\partial x} \left[ H(x,\xi) 
- A(x,\xi) \right]
\eeq
is continuous at $x = \xi$.  This important property comes about as
follows: As can be seen from the graphical representation of the reduction
formula, Fig.~\ref{fig_reduc}, $x = \xi$ corresponds to the point
where integration contour in Eq.(\ref{reduction}) passes through the upper
corner of the ``rhombus'' representing the two--dimensional region of
support of the double distribution, $f(\tilde x , \alpha)$. For a general
double distribution, which not necessarily vanishes at 
$\alpha = \pm |\tilde x|$, the SPD's $H$ and $A$ individually will be
continuous at $x = \xi$, while their first derivative in $x$ will generally
have a finite discontinuity. However, the difference, $H - A$, can be
written as an integral of the type of Eq.(\ref{reduction}) but with a
factor of $1 - \alpha$, which suppresses the contributions located in the
corner of the integration region.  As a result, not only the function 
$H - A$ but also its first derivative in $x$ are continuous at $x = \xi$.
\par
{\it Matrix elements of vector--type string operators in terms of the
transformed SPD $R_W$.}
In terms of the new function, $R_W$, we can express the matrix elements of the 
vector and axial vector operators, Eqs.(\ref{para7}) and (\ref{parax5}), as
\be 
\lefteqn{
\langle p - r/2 | \left[ \bar\psi (-z/2) \gamma_\sigma \psi (z/2) 
\right]^{\mbox{\scriptsize sym + total der.}} 
| p + r/2 \rangle } 
&& \nonumber \\
&=& \int_{-1}^1 d x  \, e^{-i x(pz) }
\left\{ 
2 p_{\sigma} 
\left[ H(x, \xi) - \frac{i (\Delta z)}{2} A(x, \xi) \right]
\right.
\nonumber \\ 
&+&  \left. \frac{1}{2} \, \left[ \Delta_{\sigma} 
- p_{\sigma}\frac{(\Delta z)}{(pz)} \right]
\left[ R_W (x, \xi) - R_W (-x, \xi)  \right] \right\} ,
\label{para8} 
\ee
\be 
\lefteqn{
\langle p - r/2 | 
\left[ \bar\psi (-z/2) \gamma_\sigma \gamma_5 \psi (z/2) 
\right]^{\mbox{\scriptsize sym + total der.}} 
| p + r/2 \rangle }
&& \nonumber \\
&=& \frac{i}{2} \, \epsilon_{\sigma\alpha\beta\gamma} \,
\frac{ z_\alpha}{(pz)} 
 \Delta_{\beta}    p_{\gamma}
\int_{-1}^1 d x \,e^{-i  x (pz) } 
 \left[ R_W (x,\xi) + R_W (-x,\xi) \right] .
\label{parax6} 
\ee
Note that only the odd part of $R_W (x, \xi)$ contributes to the
vector operator, Eq.(\ref{para8}), and only the even part to the
axial vector one, Eq.(\ref{parax6}). [We remind that the full
vector matrix element includes also the contribution from the
PW--term, Eq.(\ref{para_PW}), which remains unchanged.]
\par
We can now establish the relation of our functions to the
twist--3 SPD's introduced within the collinear expansion approach
by Anikin {\it et al.} \cite{Anikin:2000em}. In Ref.\cite{Anikin:2000em}
the matrix elements of the vector and axial vector operator were 
parametrized in the form (using our notation)
\be
\frac{1}{2} \int \frac{d\lambda}{2\pi} e^{i\lambda x} 
\langle p - r/2 | \bar\psi (-\lambda n/2) 
\left\{ \begin{array}{c} \gamma_\sigma \\[1.5ex]
\gamma_\sigma \gamma_5 \end{array}  \right\}
\psi (\lambda n /2) 
| p + r/2 \rangle 
&=& 
\left\{ \begin{array}{r} 
p_{\sigma}   H_1 (x, \xi) - \Delta_\sigma H_3 (x, \xi) ,
\\[1.5ex]
{\displaystyle \frac{i}{4}} \, \epsilon_{\sigma\alpha\beta\gamma} \,
n_\alpha \Delta_\beta p_\gamma \, H_A (x, \xi) .
\end{array} 
\right.
\label{Anikin_SPD}
\ee
where $n$ is a light--like vector, $n^2 = 0$, normalized as $(pn) = 1$, and
$(\Delta n) = 0$. Setting in Eqs.(\ref{para8}) and (\ref{parax6}) 
$z = \lambda n$, so that the terms proportional $(\Delta z)$ drop out, and
comparing with Eqs.(\ref{Anikin_SPD})
we find that the Wandzura--Wilczek--type parts of these functions are given by
\be
H_1 (x, \xi )^{\rm W.W.} &=& H(x, \xi) ,
\\
H_3 (x, \xi )^{\rm W.W.} &=& -\frac{1}{4} \left[ 
R_W (x, \xi ) - R_W (-x, \xi) \right] ,
\\
H_A (x, \xi )^{\rm W.W.} &=& \phantom{-} \frac{1}{4} \left[ 
R_W (x, \xi ) + R_W (-x, \xi) \right] .
\ee
\par
{\it Compton amplitude in terms of the transformed SPD $R_W$.}
We can also express the Compton amplitude in terms of the new
function, Eq.(\ref{rw}). This can be done either by computing the
Compton amplitude as in Section~\ref{subsec_amplitude}, using the
parametrizations Eqs.(\ref{para8}) and (\ref{parax6}) for the 
matrix elements of the vector--type operators, or directly by
rewriting the old result, Eq.(\ref{Compton_Anikin}), in terms of the
new SPD, Eq.(\ref{rw}). We obtain
\be
T_{\mu\nu} &=& \frac{1}{2(pq)} \left\{ \; 
2 \left[ p_\mu q_\nu + q_\mu p_\nu - g_{\mu\nu} (pq) + 2 \xi p_\mu p_\nu
- \frac{\Delta_\mu}{2} p_\nu  + p_\mu \frac{\Delta_\nu}{2}  
\right] \int_{-1}^1 dx \frac{H(x, \xi)}{x - \xi + i0} 
\right.
\nonumber
\\[2ex]
&& \phantom{\frac{1}{2(pq)}} + \;  (q_\mu + 3 \xi p_\mu) \Delta_\nu
\int_{-1}^1 dx \, \frac{R_W (x, \xi)}{\xi + x - i0}
\nonumber
\\[2ex]
&& \phantom{\frac{1}{2(pq)}} \left. + \; \Delta_\mu (q_\nu + \xi p_\nu)
\int_{-1}^1 dx \,  
\frac{R_W (x, \xi)}{-\xi + x + i0} \;
\right\} .
\label{comfin} 
\ee
To this one should add the contribution from the PW--term, 
Eq.(\ref{PW}), which remains unchanged.
\par
{\it Singular integrals with the transformed SPD $R_W$.} 
Let us now inspect the singular integrals in the new representation of the
Compton amplitude, Eq.(\ref{comfin}). In particular, this will provide an
alternative explanation of the divergence observed in
Section~\ref{subsec_amplitude}.  In the second term, contributing to the
amplitude for longitudinal photon polarization, the integral converges,
since $R_W (x, \xi )$ is continuous at $x = -\xi$. This follows immediately
from the integral definition of $R_W$, Eq.(\ref{rw}), and the fact that the
function $R(x, \xi)$ in the integrand is bounded at $x = -\xi$.  In the
third term, however, contributing to the amplitude for transverse photon
polarization, the integral with the singular function $1/(x - \xi +i0)$
would converge only if the function $R_W(x, \xi)$ were continuous at
$x=\xi$. However, according to the definition, Eq.(\ref{rw}), $R_W$ is
given by the integral of the function $R(y,\xi)/(y-\xi)$ from $x$ to 1 if
$x> \xi$, and from $x$ to $-1$ if $x < \xi$. Hence there is no {\em a
priori} reason why the limiting values of $R_W(x,\xi)$ when approaching
$x=\xi$ from below and from above should coincide. In fact, the difference
of the two limiting values can be written as a principal value integral:
\beq
R_W(\xi + 0, \xi) -  R_W(\xi - 0, \xi )
\;\; = \;\; {\rm P} \int_{-1}^1 dy \; \frac{R(y,\xi)}{y-\xi} ,
\label{pv}
\eeq
which is generally non-zero [in fact, this is just the real part of the
integral Eq.(\ref{Ixi})]. We thus conclude that in this term
the $x$--integral is divergent.
Thus, our conclusions regarding the existence
of the singular integrals in the terms contributing for
longitudinal and transverse polarization are the same as those
made in Section~\ref{subsec_amplitude}. This should be so; after all,
the introduction of the Wandzura--Wilczek--type SPD, $R_W$, merely 
corresponds to a change of integration variables in the integrals
in the second and third terms of Eq.(\ref{Compton_Anikin}). Only the
interpretation of the divergence is now different: It happens because
of a discontinuity of the transformed function, $R_W (x, \xi )$, 
at $x = \xi$. As we have seen, this discontinuity is built into 
the transformation at a very elementary level; it is not contingent 
on any dynamical properties of the underlying SPD's, $H$ and $A$.
This is in accordance with the fact that in the formulation 
of Section~\ref{subsec_amplitude} the divergence in the $v$--integral
is not influenced by the behavior of the SPD's.
\section{Conclusions}
\label{sec_conclusions}
In this paper we have studied the DVCS amplitude at tree level,
making use of the light--cone expansion in terms of QCD string 
operators in coordinate space. Let us summarize our main conclusions.
\par
First, we have found that transversality of the light--cone expansion can
be maintained by including in addition to the twist--2 operators a minimal
set of ``kinematical'' twist--3 operators, which appear as total
derivatives of twist--2 operators. The ``dynamical'' twist--3 contribution
from quark--gluon operators is not involved in restoring transversality of
the twist--2 contribution; the contribution from quark--gluon operators
should be separately transverse once also their total derivatives are
included (this can be done by generalizing the iterative procedure
described in Section~\ref{subsec_transversality} and 
Appendix~\ref{app_deconstructing}). From the point of view
of transversality alone it is thus a legitimate approximation to neglect
this contribution.\footnote{This situation with transversality in the
light--cone expansion is very much reminiscent of the old problem of
current conservation in the Bethe--Salpeter equation with reduced kernels,
also known under the name of ``exchange currents''
\cite{Mandelstam:1955sd,Gross:1987bu}. For a given truncation of the full
Bethe--Salpeter kernel the ``kinematical'' modifications to the current
operator, required to maintain transversality, can be computed from the
truncated kernel in a well--defined way. In addition, however, there can
appear ``dynamical'' exchange currents, which are separately
transverse. See Ref.\cite{Gross:1987bu} for a discussion of this problem.}
In this sense the original appeal of DVCS as a process describable in terms
of a small set of twist--2 non-forward matrix elements, closely related to
the parton distribution of DIS, is preserved.
\par
Second, we have found that retaining only the ``kinematical'' twist--3
contribution required by transversality leads to a finite result for the
tensor amplitude corresponding to longitudinal polarization of the initial
(virtual) photon. The tensor amplitude for transverse polarization contains
a divergence already at tree level, whose origin are contributions with
vanishing ``longitudinal'' component (in the parton model sense) of the
momentum of the quark propagator. This divergence is specific to DVCS
kinematics (technically speaking, they occur only if $\eta = \xi$) and are
absent for finite virtuality of the outgoing photon. However, we found that
the divergent part of the tensor amplitude has zero projection on the
polarization vector of the outgoing (real) photon. Thus, in both cases
(longitudinally or transversely polarized initial photon) the physical
helicity amplitudes are finite. In this sense the approximation of keeping
only the ``kinematical'' twist--3 contributions is consistent with the
factorization of the DVCS amplitude. It may thus be used for estimating
observables such as the DVCS cross section, the contribution from
interference of DVCS with the Bethe--Heitler process, spin and charge
asymmetries, {\it etc.}
\par
Our results can serve as a basis for computing kinematical power
corrections to the DVCS amplitude, both of type $t/q^2$ and $m^2/q^2$,
which are analogous to the well--known target mass corrections in DIS
\cite{Nachtmann:1973mr,Georgi:1976ve}.  In the coordinate--space approach
these corrections can be incorporated in a very elegant way, by including
terms of order $z^2$ in the parametrizations of the matrix elements of the
twist--2 operators.  The latter are obtained by solving the harmonic
condition on the matrix elements, Eq.(\ref{harmonic_O})
\cite{Balitsky:1989bk,Balitsky:1989fi}.
\par
In this paper we have computed the DVCS amplitude for the pion (more
precisely, its isoscalar component), taking advantage of the fact that the
pion has spin zero and definite $C$--parity.  The generalization of the
approach described in Section~\ref{sec_amplitude} to the more realistic
case of the nucleon is straightforward. The parametrization of the nucleon
matrix elements of the twist--2 string operators in terms of double
distributions and a PW--term has been discussed in
Refs.\cite{Polyakov:1999gs,Radyushkin:1999bz}.  In particular, this will
allow to study the influence of the ``kinematical'' twist--3 contributions
on the sum rules relating the nucleon SPD's in the forward limit to the
quark orbital angular momentum in the nucleon \cite{Ji:1997nm}.
\section{Acknowledgements}
We are grateful to I.I.~Balitsky, J.~Bl\"umlein, V.M.~Braun, M.~Diehl,
B.~Geyer, N.~Kivel, M.~Lazar, D.~M\"uller, I.~Musatov, M.V.~Polyakov, and
O.~Teryaev for many interesting discussions ad helpful hints.  While this
work was being completed the article of Belitsky and M\"uller appeared
\cite{Belitsky:2000vx}, where a similar recursive method for inclusion of
the ``kinematical'' twist--3 contributions is described. The divergences in
the amplitude for transverse photon polarization were also noted recently
by Kivel {\it et al}. \cite{Kivel:2000rb}. The generalization to the case
of a nucleon target has been discussed in Ref.\cite{Kivel:2000cn}.
\newpage
\appendix
\section{Deconstructing the string operators}
\label{app_deconstructing}
In this appendix we demonstrate how the operator relations 
Eqs.(\ref{rest_total_derivative}) and (\ref{rest_5_total_derivative}), 
expressing the vector string operator through its
symmetric part and the total derivative of the string operator
of opposite parity, can be solved recursively, giving rise to the 
closed--form expression stated in Eq.(\ref{string_deconstructed}).  
As explained in Section~\ref{subsec_transversality}, we shall not consider the 
quark--gluon operators appearing in the reduction process, since they are not 
needed to maintain transversality of the light--cone expansion. In principle, of 
course, generalizing the technique described below, the operator relations could 
be solved keeping also the quark--gluon operators.
\par
For brevity let us introduce the following notation for the 
basic (unsymmetrized) vector and axial vector string operators:
\beq
\left. \begin{array}{c} S_\sigma (z | X) \\[1.5ex]
S_{5 \sigma} (z | X) \end{array} \right\} 
\;\; \equiv \;\; 
\bar\psi (X - z/2) \left\{ \begin{array}{c} \gamma_\sigma \\[1.5ex]
\gamma_\sigma \gamma_5 \end{array}  \right\} \psi (X + z/2) .
\eeq
The R.H.S.\ of the relations Eqs.(\ref{rest_total_derivative}) and 
(\ref{rest_5_total_derivative})
can be regarded as the action of a certain linear integral operator on 
the string operators:
\beq
({\cal L} S)_\sigma (z | X) \;\; \equiv \;\; \frac{1}{2} 
\epsilon_{\sigma\alpha\beta\gamma} z_\alpha R_\beta
\int_0^1 dt\, t \; S_\gamma (tz | X) ,
\label{L_operator}
\eeq
where
\beq
R_\beta \;\; \equiv \;\; i \frac{\partial}{\partial X_\beta} .
\eeq
With the help of this formal operator, 
Eqs.(\ref{rest_total_derivative}) and (\ref{rest_5_total_derivative}),
with the quark--gluon operators neglected, can be written in compact form:
\be
- ({\cal L} S)_\sigma \; + \; S_{5\sigma} &=& S^{\rm sym}_{5\sigma} ,
\label{S_system_1}
\\
- ({\cal L} S_5 )_{\sigma} \; + \; S_\sigma &=& S^{\rm sym}_\sigma .
\label{S_system_2}
\ee
These equations decouple if one introduces instead of the vector and
axial vector type string operators the left-- and right--handed 
combinations $(S_\sigma \pm  S_{5\sigma})/2$; 
the decoupling is, of course, a consequence 
of chiral invariance of QCD. The formal solution of the system 
Eqs.(\ref{S_system_1}) and (\ref{S_system_2}) is
\be
S_\sigma &=& 
\left( \frac{1}{1 - {\cal L}^2} S^{\rm sym} \right)_\sigma
\; + \;
\left( \frac{{\cal L}}{1 - {\cal L}^2} S_5^{\rm sym} \right)_\sigma ,
\label{S_solution_L}
\\
S_{5\sigma} &=& \left( \frac{1}{1 - {\cal L}^2} S_5^{\rm sym} \right)_\sigma
\; + \; 
\left( \frac{{\cal L}}{1 - {\cal L}^2} S^{\rm sym} \right)_\sigma .
\label{S_5_solution_L}
\ee
These relations allow to express the full string operators in terms
of totally symmetrized ({\it i.e.}, twist--2) operators. The
inverse of the linear operator $1 - {\cal L}^2$ here is understood 
in the sense of a series expansion:
\be
\frac{1}{1 - {\cal L}^2} &=& 1 + {\cal L}^2 + {\cal L}^4 + \ldots
\label{L_series_even}
\\
\frac{{\cal L}}{1 - {\cal L}^2} &=& {\cal L} + {\cal L}^3 
+ {\cal L}^5 + \ldots
\label{L_series_odd}
\ee
By iterating the operator ${\cal L}$ one can easily convince
oneself that the even and odd powers of ${\cal L}$ observe the following
patterns ($n \geq 1$):
\be
({\cal L}^{2n} S)_\sigma &=& \frac{(-1)^{n-1}}{2 (2 n - 1)!} 
\left[ R_\sigma z_\gamma - (Rz) g_{\sigma\gamma} \right]
\left[ \frac{(Rz)}{2}\right]^{2n - 1}
\int_0^1 dv\; v \bar v^{2n - 1} \; S_{\gamma} (vz) ,
\label{pattern_even}
\\
({\cal L}^{2n - 1} S)_\sigma &=& \frac{(-1)^{n-1}}{2 (2 n - 2)!} 
\epsilon_{\sigma\alpha\beta\gamma} z_\alpha R_\beta
\left[ \frac{(Rz)}{2} \right]^{2n - 2}
\int_0^1 dv\; \; v \bar v^{2n - 2} \; S_{\gamma} (vz) .
\label{pattern_odd}
\ee
These expressions apply up to terms involving operators of the form 
\beq
R^2 S_\gamma  \hspace{2em}\mbox{or} \hspace{2em} R_\gamma S_\gamma ,
\label{dropped_operators}
\eeq
whose matrix elements in DVCS kinematics are proportional to $t$ or $m^2$.
More rigorously, Eqs.(\ref{pattern_even}) and (\ref{pattern_odd}) can
be proven be induction. Obviously, 
Eq.(\ref{pattern_odd}) with $n = 1$ describes the first power of the 
operator, ${\cal L}$, Eq.(\ref{L_operator}). Acting on Eq.(\ref{pattern_odd})
(for general $n$) with ${\cal L}$ we have to remember that 
Eq.(\ref{L_operator}) implies
that the distance, $z$, is to be rescaled by the parameter, $t$, everywhere,
not only in the arguments of the fields, so that we pick up
a total factor of $t^{2n -1}$. We thus obtain:
\be
[{\cal L} ({\cal L}^{2n - 1} S)]_\sigma &=& ({\cal L}^{2n} S)_\sigma
\;\; = \;\; 
\frac{(-1)^{n-1}}{4 (2 n - 2)!} \epsilon_{\sigma\alpha\beta\gamma}
\epsilon_{\gamma\alpha'\beta'\gamma'} z_\alpha z_{\alpha'} R_\beta R_{\beta'}
\nonumber \\
&& \times \left[ \frac{(Rz)}{2}\right]^{2n - 2}
\int_0^1 dt\, t^{2n} \int_0^1 du\, u \bar u^{2n - 2} \; S_{\gamma'} (tuz) .
\label{odd_even}
\ee
Using the well--known expression for the contraction of two 
$\epsilon$--tensors we find
\beq
\epsilon_{\sigma\alpha\beta\gamma}
\epsilon_{\gamma\alpha'\beta'\gamma'} z_\alpha z_{\alpha'} R_\beta R_{\beta'}
\;\; = \;\; (Rz) \left[ 
R_\sigma z_{\gamma'} - (Rz) g_{\sigma\gamma'} \right]
\; + \; \mbox{terms $\propto R^2, R_{\gamma'}$}.
\eeq
The parameter integral in Eq.(\ref{odd_even}) can be simplified by
introducing a new variable, $v \equiv tu$:
\be
&& \int_0^1 dt\, t^{2n} \int_0^1 du\, u (1 - u)^{2n - 2} \; S_{\gamma'} (tuz) 
\nonumber
\\
&=& \int_0^1 dt\, \int_0^t dv\, v (t - v)^{2n - 2} \; S_{\gamma'} (vz)
\nonumber
\\
&=& \int_0^1 dv\, \int_v^1 dt\, v (t - v)^{2n - 2} \; S_{\gamma'} (vz)
\nonumber
\\
&=& \frac{1}{2 n - 1} \int_0^1 dv\, v (1 - v)^{2n - 1} \; S_{\gamma'} (vz) .
\label{odd_even_integral}
\ee
Combining these statements we see that Eq.(\ref{odd_even}) is indeed 
equal to the proposed expression, Eq.(\ref{pattern_even}). 
In an analogous way one shows that acting on Eq.(\ref{pattern_even}) 
with ${\cal L}$ one reproduces Eq.(\ref{pattern_odd})
with $n$ replaced by $n + 1$.
\par
From Eqs.(\ref{pattern_even}) and (\ref{pattern_odd})
we see that the series of even and odd powers of the operator, ${\cal L}$,
Eqs.(\ref{L_series_even}) and (\ref{L_series_odd}), have the form of
the power series of the sine and cosine functions, respectively:
\be
\sum_{n = 1}^\infty \frac{(-1)^{n - 1}}{(2n - 1)!}
\left[ \bar v \frac{(Rz)}{2}\right]^{2n - 1} 
&=& \sin \left[ \bar v \frac{(Rz)}{2}\right] ,
\\
\sum_{n = 1}^\infty \frac{(-1)^{n - 1}}{(2n - 2)!}
\left[\bar v \frac{(Rz)}{2}\right]^{2n - 2} 
&=& \cos \left[\bar v \frac{(Rz)}{2}\right] .
\ee
Substituting this in Eqs.(\ref{S_solution_L}), (\ref{S_5_solution_L}) 
we finally obtain [{\it cf.}\ Eq.(\ref{string_deconstructed})]:
\be
S_\sigma (z) &=& S_\sigma^{\rm sym} (z) \; + \;
\frac{1}{2} \left[ R_\sigma z_\gamma 
- (Rz) g_{\sigma\gamma} \right]
\; \int_0^1 dv \, v \; \sin \left[ \bar v \frac{(Rz)}{2} \right] 
S_\gamma^{\rm sym} (vz)
\nonumber \\
&+& \frac{1}{2} \epsilon_{\sigma\alpha\beta\gamma} z_\alpha 
R_\beta \; \int_0^1 dv \, v \; \cos \left[ \bar v \frac{(Rz)}{2} \right]
S_{5\gamma}^{\rm sym} (vz) ,
\ee
and an analogous result with $S$ and $S_5$ interchanged. 
Note that this expression applies up to operators of the form
Eq.(\ref{dropped_operators}), which give power--suppressed contributions
to the DVCS amplitude.
\par
We remind again that the full expression for the string operator contains
also quark--gluon string operators, which upon symmetrization produce
operators of twist $\geq 3$. We have not kept track of them, since they
turn out not be essential for restoring transversality of the twist--2
contribution. In principle they could be included in the above iterative
procedure.
\end{document}